\documentclass[twocolumn]{aastex631}

\newcommand{\Fmag}{\textit{F146} }

\shorttitle{Transits around Mid-M Dwarfs and UCDs with Roman}
\shortauthors{Tamburo et al.}

\received{2023 March 17}

\revised{2023 April 26}

\accepted{2023 May 1}

\submitjournal{AJ}

\begin{document}

\title{Predicting the Yield of Small Transiting Exoplanets around Mid-M and Ultra-Cool Dwarfs in the Nancy Grace Roman Space Telescope Galactic Bulge Time Domain Survey}

\correspondingauthor{Patrick Tamburo}
\email{tamburop@bu.edu}

\author[0000-0003-2171-5083]{Patrick Tamburo}
\affiliation{Department of Astronomy \& The Institute for Astrophysical Research, Boston University, 725 Commonwealth Ave., Boston, MA 02215, USA}

\author[0000-0002-0638-8822]{Philip S. Muirhead}
\affiliation{Department of Astronomy \& The Institute for Astrophysical Research, Boston University, 725 Commonwealth Ave., Boston, MA 02215, USA}

\author[0000-0001-8189-0233]{Courtney D. Dressing}
\affiliation{Department of Astronomy, 501 Campbell Hall, University of California at Berkeley, Berkeley, CA 94720, USA}

\begin{abstract}
    We simulate the yield of small (0.5$\textendash$4.0~R$_\oplus$) transiting exoplanets around single mid-M and ultra-cool dwarfs (UCDs) in the Nancy Grace Roman Space Telescope Galactic Bulge Time Domain Survey. We consider multiple approaches for simulating M3$\textendash$T9 sources within the survey fields, including scaling local space densities and using Galactic stellar population synthesis models. These approaches independently predict $\sim$100,000 single mid-M dwarfs and UCDs brighter than a Roman \textit{F146} magnitude of 21 that are within the survey fields. Assuming planet occurrence statistics previously measured for early-to-mid M dwarfs, we predict that the survey will discover 1347$^{+208}_{-124}$ small transiting planets around these sources, each to a significance of 7.1$\sigma$ or greater. Significant departures from this prediction would test whether the occurrence rates of small planets increase or decrease around mid-M dwarfs and UCDs compared to early-M dwarfs. We predict the detection of 13$^{+4}_{-3}$ habitable, terrestrial planets ($R_p$~$<$~1.23~R$_\oplus$) in the survey. However, atmospheric characterization of these planets will be challenging with current or near-future space telescope facilities due to the faintness of the host stars. Nevertheless, accurate statistics for the occurrence of small planets around mid-M dwarfs and UCDs will enable direct tests of predictions from planet formation theories and will determine our understanding of planet demographics around the objects at the bottom of the main sequence. This understanding is critical given the prevalence of such objects in our Galaxy, whose planets may therefore comprise the bulk of the galactic census of exoplanets.
    
\end{abstract}

\section{Introduction}
\label{sec:intro}

The NASA Kepler mission \citep{Borucki2010} revealed a significant anti-correlation between stellar host mass and the occurrence rates of short-period, small exoplanets \citep[$P~\lesssim~200$~days, $R_p$~$\leq$~4.0~$R_\oplus$;][]{Howard2012,Mulders2015a}. Early-M dwarfs were found to host short-period super-Earths ($1.0$~$R_\oplus$~$<$~$R_p$~$<$~$2.8$~$R_\oplus$) at a rate roughly three times higher than that of F dwarfs in the Kepler field \citep{Mulders2015a}, and possess around 2.5 short-period planets with  $R_p \leq 4.0$~$R_\oplus$ per star, on average \citep{Dressing2013,Dressing2015,Gaidos2016}. This anti-correlation has also been observed in radial velocity (RV) surveys \citep{Sabotta2021,Pinamonti2022}.


Naively, the anti-correlation between host mass and short-period small planet occurrence rates is surprising, since lower-mass stars have, on average, less massive reservoirs of dust in their protoplanetary disks with which to form planets through core accretion \citep{Andrews2013,Pascucci2016}. Factors beyond disk mass alone must be at play to explain the observed trends. One possibility is that giant planets, which occur more frequently around higher-mass hosts \citep[e.g.,][]{Johnson2010}, limit the flux of planet-forming pebbles to the inner disk and hence preferentially suppress the formation of short-period super-Earths around higher-mass stars. \cite{Mulders2021} found that that this pebble-flux-limiting scenario can explain observed Kepler planet occurrence rates and predicted an overturn in short-period super-Earth planet occurrence rates at host masses below $\sim$0.5~$M_\odot$, with essentially no super-Earths formed at an orbital distance of 0.3~au around hosts with masses below 0.1~$M_\odot$. 


However, there are currently few observational constraints on small planet occurrence rates around hosts with masses below 0.3~$M_\odot$ (SpT $\sim$M3), owing to their intrinsic faintness and spectral energy distributions that peak in the near-infrared (NIR). \citet{HardegreeUllman2019} used Kepler detections to measure the occurrence rates of planets on 0.5$\textendash$10 day orbits with radii between $0.5\textendash2.5$ R$_{\oplus}$ around mid-type M dwarfs (M3$\textendash$M5) and reported a tentative increase from $0.86^{+1.32}_{-0.68}$ to $3.07^{+5.49}_{-2.49}$ planets per star over those spectral types; however, their small sample size of 13 planets around 7 stars lead to occurrence rates with large fractional uncertainties. Their results are supported by occurrence rates measured with RV detections from the CARMENES survey, which found a significant increase in the occurrence rate of low-mass ($1$~$M_\oplus$~$<$~$M_p\sin{i}$~$<$~$10$~$M_\oplus$) planets with periods less than 10 days around stars with $M_\star$~$<$~$0.34$~$M_\odot$ compared to higher mass stars \citep{Sabotta2021}. However, recent results from the Transiting Exoplanet Survey Satellite \citep[TESS;][]{Ricker2015} suggest that the mission's yield of $0.5\textendash2.0$~$R_\oplus$ planet candidates around nearby mid-M dwarfs is best produced by a constant or decreasing planet occurrence rate compared to early-M dwarfs \citep{Brady2022,Ment2023}.


Occurrence rates of small planets around ultra-cool dwarfs \citep[UCDs, spectral types M7 and later;][]{Kirkpatrick1997} are even less constrained. To date, only one system of small exoplanets has been confirmed transiting a UCD host: the seven Earth-sized planets around TRAPPIST-1 \citep{Gillon2016,Gillon2017}. Data from the full TRAPPIST-Ultra-Cool Dwarf Transit Survey \citep{Gillon2013a} were used to place a lower limit of 10\% on the occurrence rates of TRAPPIST-1b-like planets around their sample of 40 UCDs \citep{Lienhard2020}. Other data sets have been used to place upper limits on planet occurrence rates around UCDs.  \citet{He2017} used Spitzer observations of 44 brown dwarfs and found that on periods less than 1.28 days, the occurrence rate of planets with radii between 0.75$\textendash$3.25~R$_\oplus$ is less than 67$\pm$1\%. \citet{Sagear2020} and \citet{Sestovic2020} both used K2 data to search for transiting planets around UCDs, but were not able to strongly constrain the occurrence rates of small exoplanets due to low signal-to-noise, owing to the operation at visible wavelengths. Dedicated ground-based transit surveys at NIR wavelengths like the Search for habitable Planets EClipsing ULtra-cOOl Stars \citep[SPECULOOS;][]{Delrez2018,Murray2020,Sebastian2021}, the ExoEarth Discovery and Exploration Network (EDEN) \citep{Gibbs2020}, and the Perkins INfrared Exosatellite Survey \citep[PINES;][]{Tamburo2022a} are actively searching for small planets around UCDs, but they would have to detect hundreds of planets to determine occurrence rates to a similar precision as measured for early-M dwarfs with Kepler. 

The Nancy Grace Roman Space Telescope\footnote{Formerly the Wide Field Infrared Survey Telescope\\(WFIRST).}, currently slated for launch in 2026, is the next flagship mission of the NASA Astrophysics Division \citep{Spergel2015, Akeson2019}. It is a 2.4-m telescope that will be stationed at the second Sun-Earth Lagrange point (L2), and it will be equipped with two instruments: 1) the Coronagraph Instrument \citep[CGI;][]{Noeker2016,Mennesson2020}, a technology demonstration instrument that will perform optical imaging, polarimetry, and spectroscopy of circumstellar disks and exoplanets, and 2) the Wide Field Instrument \citep[WFI;][]{Domber2019,Domber2022}, an NIR camera for imaging and slitless spectroscopy which will be used to perform the mission's three ``Core Community Surveys", these being the High Latitude Time Domain Survey, the High Latitude Wide Area Survey, and the Galactic Bulge Time Domain Survey\footnote{\url{https://roman.gsfc.nasa.gov/observations.html}}. WFI will use a mosaic of 18 H4RG detectors to capture images with more than 200 times the field of view of the NIR channel of Hubble's WFC3 instrument, but with similar image quality \citep{Akeson2019,Domber2022}. 

The design of the Galactic Bulge Time Domain Survey (hereafter ``the survey'')  was motivated by the detection of microlensing planets \citep[see, e.g.,][]{Spergel2015,Penny2019,Johnson2020}. The microlensing technique probes a different parameter space than the transit method, being most sensitive to detecting planets at $\gtrsim$1~AU from the host star. For typical microlensing events in the Galactic Bulge, for example, the method is most sensitive to planets at a distance of $\sim$[2\textendash 4]~AU~(M/M$_\odot$)$^{1/2}$ \citep[][]{Gaudi2012,Penny2019}. A large-scale microlensing survey would, therefore, complement the known population of close-in planets from transit surveys like Kepler and TESS, as well as the populations discovered by the radial velocity and direct imaging methods. To maximize the number of microlensing detections, the survey will perform time series photometry of seven fields that make up a $\sim$2~deg$^2$ region of the Galactic bulge, using NIR filters to reduce the effect of dust extinction in the Galactic plane. These seven fields will be targeted in six 72-day seasons over the course of a nominal five-year primary mission, with photometry performed using an \Fmag filter ($0.93\textendash2.00$~$\mu$m) and an \textit{F087} filter ($0.76\textendash0.98$~$\mu$m). The survey will primarily be executed in the \Fmag filter, with one $46.8$-s \Fmag exposure obtained in each of the seven fields every 909.6 seconds. The secondary \textit{F087} filter will be used to obtain color information about the sources, with one $286$-s \textit{F087} exposure taken every $12$ hours. We refer readers to \citet{Penny2019} for an in-depth summary of the current ``Cycle 7" design of the survey, but note that its exact implementation (e.g., number of fields, cadence, filters, etc.) is still being refined.

\citet{Spergel2015} estimated that the survey will provide photometry with a precision of 1\% or better for around 20 million dwarf stars. The large number of target stars, high photometric precision, and the 15-minute cadence will allow the survey to double as a search for transiting exoplanets; the NIR wavelength coverage makes it particularly well-suited to search for transiting planets around mid-M dwarfs and UCDs. \citet{Montet2017} simulated the transiting planet yield of the survey, estimating that more than $100,000$ could be detected around FGKM stars, including several thousand around M dwarfs, specifically. However, they only considered planets as small as 2 $R_\oplus$, which neglects the sizable population of 0.5$\textendash$2.0 $R_\oplus$ planets that have been detected around early-M dwarf stars \citep{Dressing2015, Gaidos2016}. They also did not consider the detection of planets around L- and T-type dwarfs, spectral types that may host an abundance of short-period planets \citep[e.g.,][]{Limbach2021,Tamburo2022b}. Recent work by \citet{Limbach2022} on the proposed Transiting Exosatellites, Moons, and Planets in Orion (TEMPO) Survey has explored the possibility of applying Roman Guest Observer time series observations to detecting planets around low-mass stars, brown dwarfs, and free-floating planets in the Orion Nebular Cluster (ONC). They estimate that a 30 day survey of the ONC could detect 14 transiting satellites around free-floating planets and 54 transiting satellites around brown dwarfs, demonstrating the power of the observatory for detecting transiting companions around hosts at the bottom of the main sequence and beyond.

In this paper, we present a simulation of the transiting planet yield around mid-M spectral types and later in the Core Community Galactic Bulge Time Domain Survey, specifically, and investigate the survey's potential for extending measurements of planet occurrence rates into the ultra-cool regime. The paper is organized as follows. In Section~\ref{sec:host_simulation}, we describe the creation of a synthetic sample of mid-M dwarfs and UCDs in the Roman microlensing fields using measured space densities from nearby volume-complete samples, which is compared against results from Galactic stellar population synthesis models. It also details the time series photometry that was simulated for these targets and which was injected with a population of transiting planets. Section~\ref{sec:injection_recovery} describes the recovery of these planets, and Section~\ref{sec:results} discusses the potential implications of these discoveries.

\section{Host Simulation}
\label{sec:host_simulation}

\subsection{Simulated Photometry}
\label{subsec:photometry}

\begin{figure}
    \centering
    \includegraphics[width=\columnwidth]{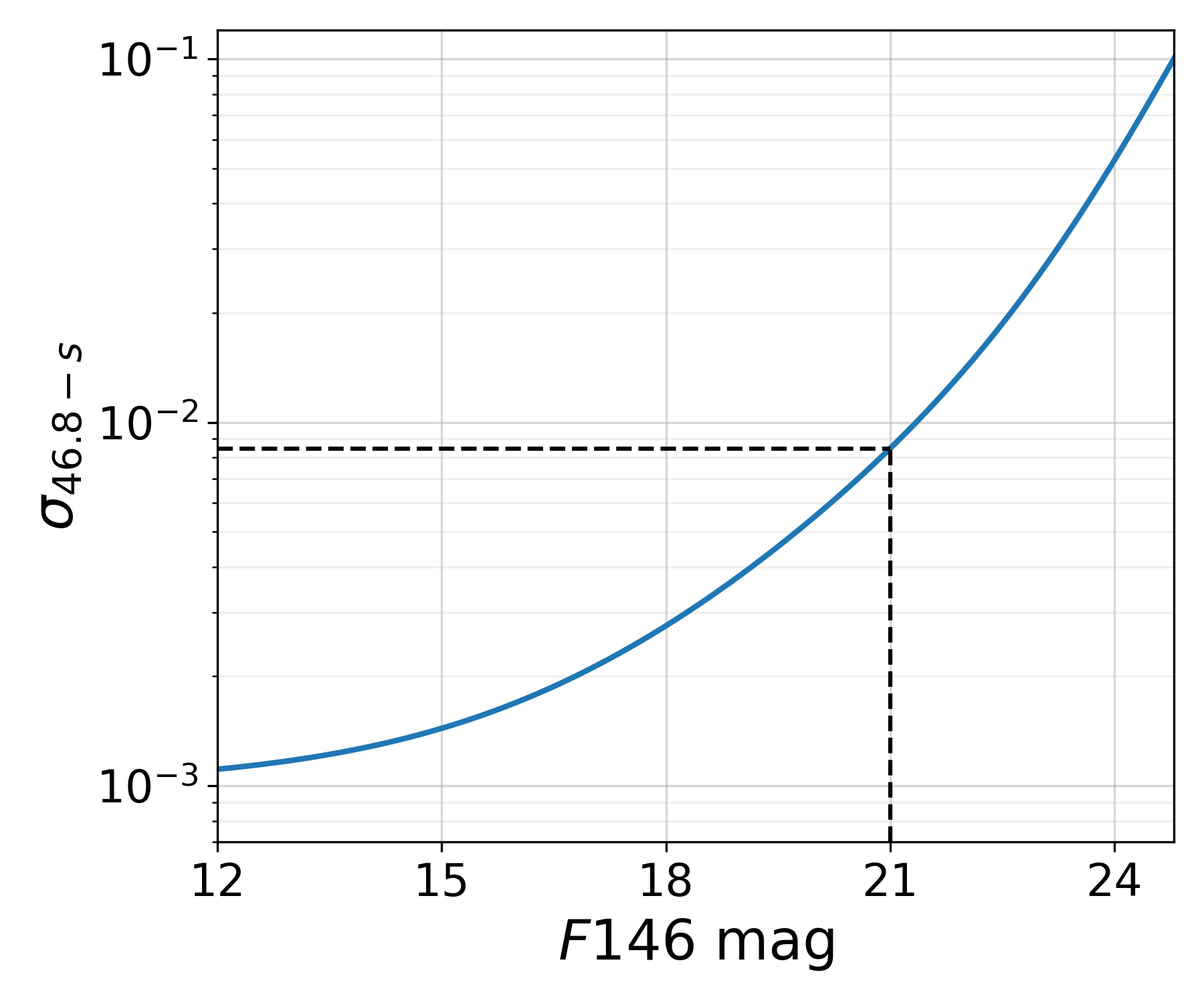}
    \caption{The standard deviation of normalized photometry in a single 46.8-s exposure in \textit{F146}-band, modeled using the parameters listed in Table \ref{tab:photometry}. The \textit{F146} = 21 magnitude cutoff and its corresponding $\sim$1\% single-exposure standard deviation are indicated in dashed black lines.}
    \label{fig:noise}
\end{figure}

The expected noise performance of the WFI drives the magnitude and distance limits that we apply in our simulations. The time domain survey will primarily use a \Fmag filter. This filter will provide wavelength coverage from 0.93\textendash 2.00~$\mu$m, entirely encompassing the standard \textit{Y}, \textit{J}, and \textit{H} photometric passbands \citep[e.g.,][]{Bessell2005}. We modeled the standard deviation of normalized photometry in the \Fmag filter using the properties listed in Table~\ref{tab:photometry}, the values of which were sourced from the WFIRST Cycle 7 design described in \citet{Penny2019}. The model assumes a sky background count rate that was determined using a time-dependent zodiacal light model evaluated at the midpoint of the survey seasons. It also assumes a 9-pixel aperture \citep[which was found to produce the best photometry in the crowded microlensing fields in ][]{Penny2019} and a constant systematic noise floor of 1~mmag. The assumed noise floor becomes the dominant noise component for magnitudes less than \Fmag~$\approx$~16.8, and its effect can be seen in the gradual tapering off of the curve in Figure~\ref{fig:noise} for the brightest source magnitudes. This assumption has little impact on the total planet yield of our simulations, since very few detections are made around sources brighter than \Fmag~=~16.8 (see Section~\ref{subsec:yield}).

\begin{deluxetable}{lrl}
        \tablecaption{Parameters describing the expected noise performance of Roman's Wide Field Instrument and the format of the Galactic Bulge time domain survey. Values follow the Cycle 7 survey design detailed in \citet{Penny2019}.}
        \label{tab:photometry}
        \tablehead{
            \colhead{Parameter}       & \colhead{Value} & \colhead{Units}
            }
        \startdata
           Dark current & 1.072 & e$^-$ s$^{-1}$ pix$^{-1}$\\
           Read noise & 12.12 & e$^-$ rms\\
           Sky background & 3.43 & e$^-$ s$^{-1}$ pix$^{-1}$\\
           Aperture size & 9 & pix\\
           \Fmag zero point & 2.078e-10 & erg s$^{-1}$ cm$^{-2}$ \AA$^{-1}$\\
           Noise floor & 1 & mmag\\
           Exposure time & 46.8 & s\\
           Cadence & 909.6 & s\\
           Season length & 72 & days\\
           Number of seasons & 6 & ---\\
        \enddata
    
    \tablecomments{The \Fmag zero point was sourced from the Spanish Virtual Observatory Filter Profile Service.}
\end{deluxetable}

The resulting noise model for a 46.8-s exposure as a function of \Fmag magnitude is shown in Figure~\ref{fig:noise}. A per-exposure uncertainty of 1\% is achieved around a magnitude of 21 in the \Fmag band, and we use this as the magnitude cutoff for sources throughout the remainder of this study. We note that this is the same magnitude cutoff used in \citet{Montet2017}. For the earliest spectral types considered in this study (M3), the magnitude cutoff corresponds to a maximum distance of about 3.5 kpc. See Table \ref{tab:densities} for estimates of the maximum distances of mid-M dwarfs and UCDs with \textit{F146}~$<$~21 in different effective temperature ($T_{eff}$) bins.

Data timestamps in \textit{F146}-band were generated assuming a 909.6-s cadence over six 72-day seasons. These six seasons were assumed to take place in two campaigns, with three seasons near the start of the mission and three seasons near the end \citep{Spergel2015}. The three seasons in each campaign were separated by half a year. Every 12 hours, we removed a simulated exposure in \textit{F146}-band to replicate expected interruptions caused by  photometry in the time domain survey's secondary \textit{F087} bandpass \citep{Penny2019}. While we included these interruptions, we did not simulate photometry in \textit{F087}-band because the 12-hour cadence in this filter is unlikely to sample many in-transit points for the average $\sim$1-hour transit durations around mid-M and UCD spectral types. For example, a planet on a 7.2-day period (near the median detected orbital period in our simulations, see Section \ref{subsec:yield}) with a 1-hour transit duration would be sampled by a maximum of just $\sim$5 in-transit exposures in \textit{F087}-band over the 432-day survey, compared to $\sim$200 in-transit exposures in \textit{F146}-band.

In the absence of constraints on spacecraft/detector systematics, we simulated target photometry at these time stamps using purely uncorrelated Gaussian (``white") noise based on the noise model shown in Figure \ref{fig:noise}. However, we note that systematic ``red" noise would necessarily impose a higher detection threshold to achieve the same number of statistical false positives as expected for the uncorrelated noise case considered here \citep[e.g.,][]{Pont2006}, which would in turn decrease our yield estimates.

\subsection{Simulated Host Population using Local Space Densities}
\label{subsec:hosts}
The population of mid-M dwarfs and UCDs with \textit{F146}~$<$~21 in the survey fields is not currently known. The faintest such sources have Gaia magnitudes of $\sim$26, which is well beyond the detection limits of that survey, especially in the crowded Galactic plane \citep[e.g.,][]{GaiaCollab2016,GaiaCollab2021b}. They also lie beyond the detection limits of previous NIR imaging campaigns with the Two Micron All Sky Survey  \citep[2MASS;][]{Skrutskie2006}, the Spitzer Space Telescope \citep{Churchwell2009}, the Wide-Field Infrared Survey Explorer \citep[WISE;][]{Wright2010}, the Panoramic Survey Telescope and Rapid Response System \citep[Pan-STARRS1;][]{Chambers2016}, and the DECam Plane Survey \citep[DECaPS;][]{Schlafly2018}.

In this section, we describe a simulation that we performed to estimate the number of mid-M dwarf and UCD sources in the survey fields with \textit{F146}~$<$~21 by using measured space densities from volume-complete samples in the solar neighborhood. In particular, we used the 15-pc sample of mid-M dwarfs from \citet{Winters2021} (hereafter W21) and the 20-pc sample of L, T, and Y dwarfs from \citet{Kirkpatrick2021} (hereafter K21). We elected to use only the sources in these samples that are presumed to be single objects, excluding sources in multiple systems. We did this because multiple systems, even though they may be resolved in the local samples, are unlikely to be resolved at the distances under consideration in this work (out to 3.5 kpc), and unresolved multiple systems complicate the interpretation of any detected transiting planet signals \citep[e.g.,][]{Bouma2018}. Out of the 512 M dwarfs with $0.1\textendash0.3$~M$_\odot$ in the 15 pc sample from W21, 290 are presumed to be single; out of the 525 L, T, and Y dwarfs in the 20 pc sample from K21, 438 are presumed to be single.

The W21 sample reports mass measurements for their sources, while the K21 sample reports source $T_{eff}$. For the purposes of our simulation, we wish to employ the space densities of mid-M dwarfs and UCDs in terms of one of these variables, and we chose to use $T_{eff}$. We converted the 290 mass measurements of single M dwarfs from W21 to estimates of $T_{eff}$ using the 10 Gyr isochrone from \citet{Baraffe2015}. We then counted the number of sources in the W21 sample in 150-K-wide $T_{eff}$ bins, enforcing an upper $T_{eff}$ limit of 3300 K ($\sim$M3) and a lower limit of 2850 K ($\sim$M6). We assumed Poisson errors on these counts, and converted the counts to space densities using the 15-pc sample volume. 

\begin{figure*}
    \centering
    \includegraphics[width=\textwidth]{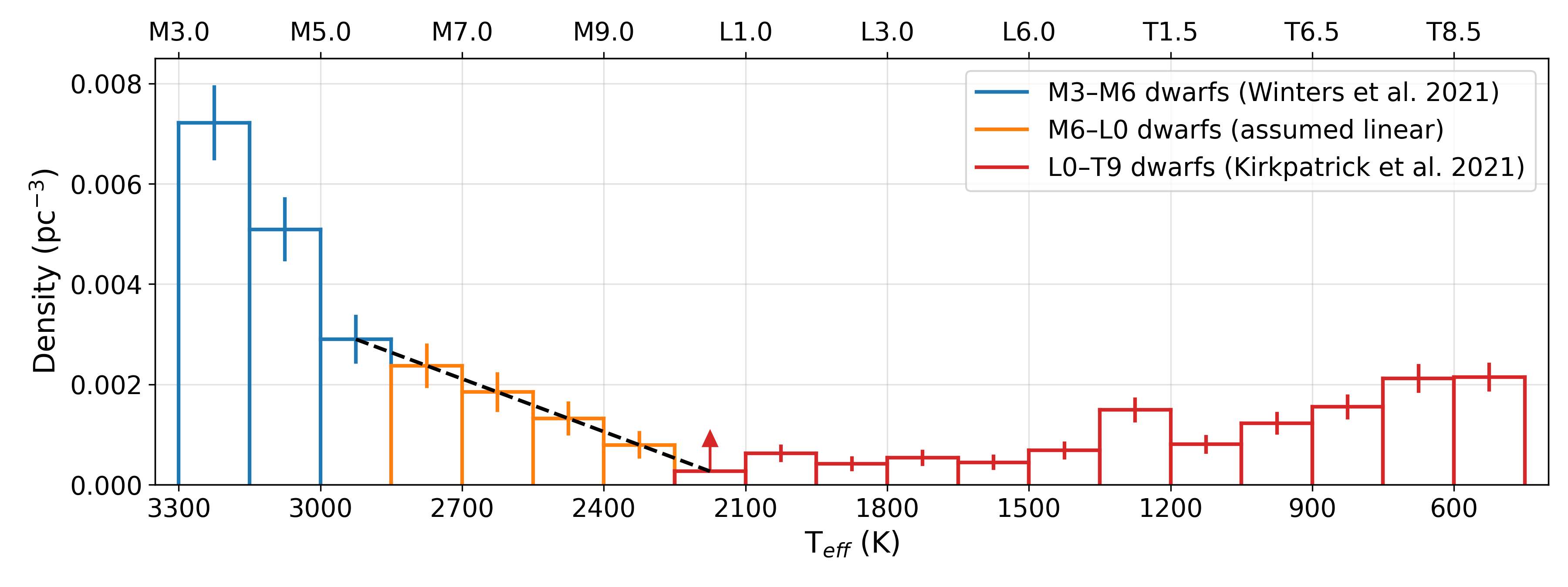}
    \caption{Local space densities of single M3\textendash T9 dwarfs as a function of effective temperature used in the simulation described in Section \ref{subsec:hosts}. The top axis is given as a function of spectral type, the values of which were determined using the M dwarf temperature scale of \citet{Rajpurohit2013} for temperatures above 2800 K and the temperature scale for field M6$\textendash$T9 dwarfs of \citet{Faherty2016} for temperatures below 2800 K. W21 densities are shown in blue and have been converted from a mass scale to a temperature scale using the 10 Gyr isochrone of \citet{Baraffe2015}. K21 densities are shown in red. The densities of M6$\textendash$M9.5 dwarfs are indicated in orange and were assumed to linearly connect the W21 and K21 densities. The linear fit that was used to generate these estimates is shown as a black dashed line. Error bars indicate the 1-$\sigma$ uncertainty on each bin, which were taken to be Poissonian. The 2100$\textendash$2250 K bin is a lower limit and is indicated with an upward arrow.}
    \label{fig:space_densities}
\end{figure*}

\begin{figure*}
    \centering
    \includegraphics[width=\textwidth]{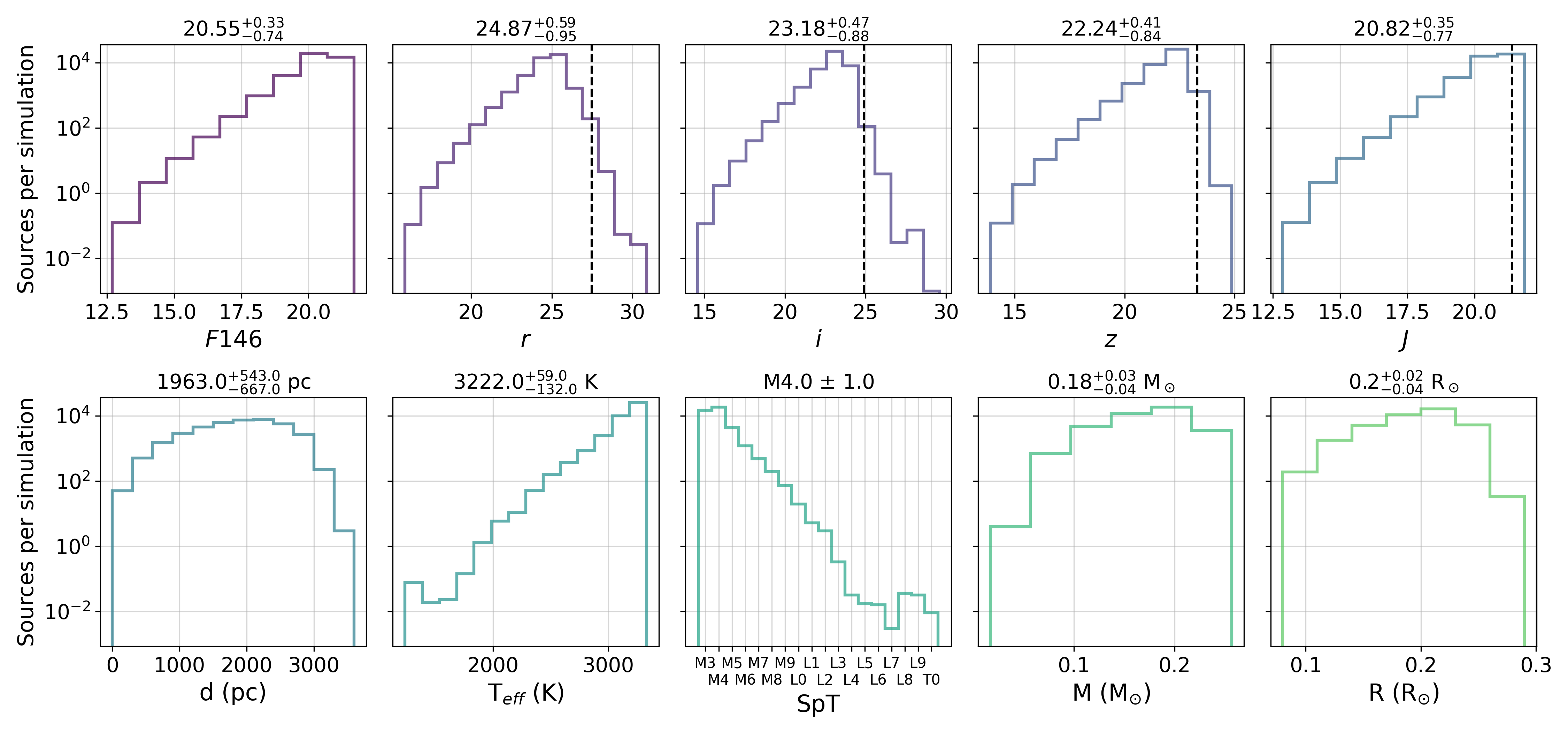}
    \caption{Distributions of host properties. The histograms are shown on a log scale and have been normalized to show the average counts in a single simulation. The 15.9\%, 50\%, and 84.1\% percentiles of each distribution (i.e., the 1$\sigma$ range) are reported at the top of each panel. Top row: source magnitudes in the \textit{F146}-, \textit{r}-, \textit{i}-, \textit{z}-, and \textit{J}-bands. Magnitudes were calculated using BT-Settl model spectra and were reduced by realistic dust extinction values. The 99.9\% percentiles for the \textit{r}, \textit{i}, \textit{z}, and \textit{J} magnitudes (used in the Besançon simulation described in Section \ref{subsubsec:besancon}) are indicated in dashed black lines. Bottom row: distance in parsecs, effective temperature in Kelvin, spectral types, mass in solar masses, and radius in solar radii.}
    \label{fig:source_histograms}
\end{figure*}

The W21 and K21 measurements do not provide space densities for $\sim$M6$\textendash$M9.5 dwarfs. Rather than exclude these types from the simulation, we filled in the gap by assuming a linear transition between the two sets of densities\footnote{The assumption of a different form for the transition (e.g., an exponential decrease) would not strongly impact our results, as M6\textendash M9.5 dwarfs make up only $\sim$5\% of all sources in our simulations with \textit{F146}~$<$~21 (see Figure \ref{fig:source_histograms}).}. This approach qualitatively reproduces the sharp decrease in source counts over the M7$\textendash$M9.5 spectral range seen in a magnitude-limited sample of 34,000 such targets in \citet{Ahmed2019}. The resulting space densities used in this study are given in Table \ref{tab:densities} and visualized in Figure~\ref{fig:space_densities}. The W21 densities are shown in blue, the K21 densities are shown in red, and the assumed densities of M6$\textendash$M9.5 dwarfs are shown in orange. The space densities are highest around M3$\textendash$M6 spectral types, reflecting the peak of the stellar initial mass function (IMF) near 0.2$\textendash$0.3~M$_\odot$ \citep[e.g.,][]{Kroupa2002,Chabrier2003,Bastian2010}. The densities then shallow out and show a rise at the latest spectral types, reflecting the fact that as sub-stellar objects age, they progressively cool and pile up in the lowest temperature bins. 

\begin{table}[]
    \centering
    \caption{Local space densities of single M3\textendash T9 dwarfs used in the simulation described in Section \ref{subsec:hosts}. Spectral types corresponding to these temperature ranges were estimated using the M-dwarf $T_{eff}$ scale from \citet{Rajpurohit2013} for temperatures above 2800 K and the field M6$\textendash$T9 $T_{eff}$ scale from \citet{Faherty2016} for temperatures below 2800 K. We also list $d_{max}$, an estimate of the maximum distance of objects within each $T_{eff}$ range with $F146$~$<$~21.} 
    \begin{tabular}{rlrr}
    \hline
    \hline
       T$_{eff}$ (K) & SpT & Density (pc$^{-3}$) & $d_{max}$ (pc)\\
       \hline
        3300\textendash3150 & M3.0\textendash M4.0 & 0.0072 $\pm$ 0.0007 & 3500\\  
        3150\textendash3000 & M4.0\textendash M5.0 & 0.0051 $\pm$ 0.0006 & 2500\\    
        3000\textendash2850 & M5.0\textendash M6.0 & 0.0029 $\pm$ 0.0005 & 1850\\
        2850\textendash2700 & M6.0\textendash M7.0 & 0.0024 $\pm$ 0.0004 & 1420\\
        2700\textendash2550 & M7.0\textendash M8.0 & 0.0018 $\pm$ 0.0004 & 1210\\
        2550\textendash2400 & M8.0\textendash M9.0 & 0.0013 $\pm$ 0.0003 & 1070\\
        2400\textendash2250 & M9.0\textendash L0.0 & 0.0008 $\pm$ 0.0002 & 840\\ 
        2250\textendash2100 & L0.0\textendash L1.0 & $\ge$0.0003 & 680\\
        2100\textendash1950 & L1.0\textendash L2.0 & 0.0006 $\pm$ 0.0001 & 620\\
        1950\textendash1800 & L2.0\textendash L3.0 & 0.0004 $\pm$ 0.0001 & 550\\
        1800\textendash1650 & L3.0\textendash L4.5 & 0.0005 $\pm$ 0.0001 & 380\\
        1650\textendash1500 & L4.5\textendash L6.0 & 0.0004 $\pm$ 0.0001 & 320\\
        1500\textendash1350 & L6.0\textendash L8.0 & 0.0007 $\pm$ 0.0001 & 270\\
        1350\textendash1200 & L8.0\textendash T1.5 & 0.0015 $\pm$ 0.0002 & 250\\
        1200\textendash1050 & T1.5\textendash T5.0 & 0.0008 $\pm$ 0.0002 & 190\\
        1050\textendash900  & T5.0\textendash T6.5 & 0.0012 $\pm$ 0.0002 & 130\\
        900 \textendash750  & T6.5\textendash T7.5 & 0.0016 $\pm$ 0.0002 & 110\\
        750\textendash600   & T7.5\textendash T8.5 & 0.0021 $\pm$ 0.0003 & 80\\
        600\textendash450  &  T8.5\textendash T9.0 & 0.0021 $\pm$ 0.0003 & 40\\
        \hline
    \end{tabular}
    \label{tab:densities}
\end{table}

\begin{figure*}
    \centering
    \includegraphics[width=\textwidth]{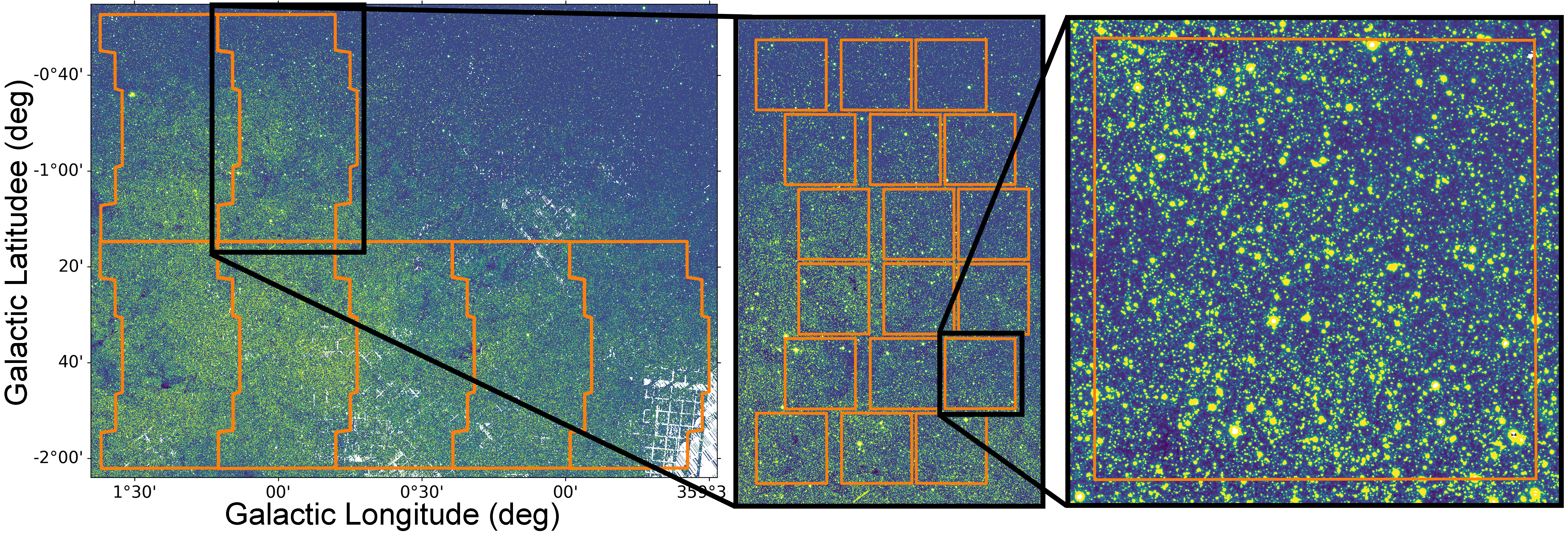}
    \caption{The on-sky footprint of the Roman Galactic Bulge Time Domain Survey with archival $z$-band imagery from Pan-STARRS1 \citep{Chambers2016}. Bad pixels in the Pan-STARRS1 data are colored in white and are used to flag saturated sources, areas without survey coverage, etc. Left: the full survey footprint, covering $\sim$2 deg$^2$ of the Galactic bulge over seven fields. Field outlines are shown in orange. Middle: a single field, with outlines of the 18 detector pointings shown in orange. Right: one detector within this field. Field pointings were taken from \citet[][\url{https://github.com/mtpenny/wfirst-ml-figures}]{Penny2019}.}
    \label{fig:roman_ps1}
\end{figure*}

The densities were measured over volumes near the Sun and do not reflect the exponential increase in star counts in the Milky Way's thin and thick disks in the direction of Galactic center. This exponential increase is commonly modeled as a function of radial distance $d$ with scale length $\ell_d$ and vertical distance above the Galactic plane $z$ with scale height $z_d$ \citep[e.g.,][]{McMillan2017}. Since the survey is performed near the Galactic midplane (see Figure \ref{fig:roman_ps1}), we ignore any scaling effects due to $z$. However, we do scale the local space densities of our simulated sources with $d$ using the following relations: 

\begin{equation}
    \label{eq:thin}
    n_t(d) = n_0\textrm{exp}(\frac{d}{\ell_t})
\end{equation}

\begin{equation}
    \label{eq:thick}
    n_T(d) = f_p n_0\textrm{exp}(\frac{d}{\ell_T})
\end{equation}.

Here, $n_0$ are the local space densities of mid-M dwarfs and UCDs given in Table \ref{tab:densities}, $\ell_t$ is the radial scale length of the thin disk, $f_p$ is the local density normalization of the thick disk to the thin disk, and $\ell_T$ is the radial scale length of the thick disk. Accurate determinations of $\ell_t$ and $\ell_T$ are notoriously difficult, as their measurement must contend with source confusion and dust extinction in the Galactic plane. A review of 130 papers on the subject by \citet{BlandHawthorn2016} revealed a range of 1.8$\textendash$6.0 kpc, for example. Measurements of $f_p$ are also difficult because it is degenerate with the scale heights of both the thin and thick disks \citep[e.g.,][]{Reyle2001}. We decided to account for this uncertainty in our simulations by drawing $\ell_t$, $\ell_T$, and $f_p$ from Gaussian distributions $N(\mu, \sigma)$, with central values $\mu$ and standard deviations $\sigma$ given by the recommendations from \citet{BlandHawthorn2016}. Each iteration, we drew $\ell_t$ from $N(2500, 400)$ pc, $\ell_T$ from $N(2000, 200)$ pc, and $f_p$ from $N(0.04, 0.02)$, where we required $f_p > 0$.


We implemented our source simulation as follows. We first divided the 3.5 kpc simulation volume up into 25 pc spherical shells\footnote{We tested the simulation with finer resolution shells and found that they did not significantly alter our results.}. In each shell, we set the space densities for sources in the temperature bins by drawing from Gaussian distributions with means and standard deviations given by the values and uncertainties reported in Table~\ref{tab:densities}. This was not done for the bin without uncertainties, which was conservatively assigned its reported lower limit each loop. We scaled these densities with the distance to the front of the shell $d$ using Equations~\ref{eq:thin} and \ref{eq:thick} to simulate populations for the thin and thick disks respectively. 

We then simulated source counts in the temperature bins within each shell by multiplying the space densities of the shell in question by the shell's volume. We multiplied these counts by the fractional area of the Roman time domain survey on the sky ($1.96$ deg$^{2}$/$41253$ deg$^{2}$) to approximate the number of sources expected within the survey footprint. We randomly positioned the resulting number of expected sources in each temperature bin across the 126 detector positions that make up the survey footprint (18 detectors at 7 field pointings; see Figure \ref{fig:roman_ps1}), assigning them random temperatures within the appropriate temperature bin and random distances within the bounds of the shell in question. Each target was also assigned a random age using a uniform distribution from $0.1\textendash10$ Gyr. 

We then assigned each target a mass $M$ and a radius $R$ given the $T_{eff}$ and age values using evolutionary models. If a source had a $T_{eff} > 2250$ K, we used the low-mass star evolutionary models of \citet{Baraffe2015}, whereas if the source had a $T_{eff} < 2250$ K, we used the Sonora Bobcat brown dwarf evolutionary models from \citet{Marley2021}. We first identified the four model points that most closely bracketed the target's $T_{eff}$ and age values. We then performed a linear 2D interpolation of the mass and radius values belonging to these four entries and used the resulting functions to evaluate the expected $M$ and $R$ for the given $T_{eff}$/age values. We used the $M$ and $R$ values to calculate the log of the surface gravity ($\log{g}$) in $cm$ $s^{-2}$.

Next, we calculated the magnitude of each source in a number of different photometric bands using BT-Settl models \citep{Allard2012} covering a temperature range of $500\textendash3300$ K (in steps of $100$ K) and $\log(g)$ values of 4.5, 5.0, and 5.5. We assumed a solar metallicity for each source. We assigned each source the closest matching model based on its $T_{eff}$ and $\log{g}$ values, scaled the model fluxes to the flux received at Earth by multiplying by a factor of $(R/d)^2$, and multiplied by the light collecting area of the telescope, assuming a primary diameter of $2.36$-m and an obscured fraction of $13.9\%$ \citep{Penny2019}. We used publicly available filter bandpasses to calculate magnitudes in Pan-STARRS \textit{r}-, \textit{i}-, and \textit{z}-bands, 2MASS \textit{J}-band, and Roman \textit{F146}-band\footnote{Pan-STARRS filters were obtained from \url{http://svo2.cab.inta-csic.es/svo/theory/fps3/index.php?mode=browse&gname=PAN-STARRS&asttype=}.\\ 2MASS filter curves were obtained from \url{http://svo2.cab.inta-csic.es/theory/fps/index.php?id=2MASS/}.\\Roman filters were obtained from \url{https://roman.gsfc.nasa.gov/science/WFI_technical.html}.}.

Finally, we extincted the magnitude in each band using the latest version of the Bayestar dust map \citep{Green2015, Green2018a, Green2019}, accessed through the Python package \texttt{dustmaps} \citep{Green2018b}. This map provides probabilistic 3D reddening estimates based on parallax measurements from Gaia DR2 and stellar colors from Pan-STARRS1 and 2MASS photometry, which can be converted to extinction estimates in various filters. We used these conversion factors to translate the reddening value returned for the 3D coordinates of each source to extinction in the \textit{r}-, \textit{i}-, \textit{z}-, \textit{J}-, and \textit{F146}-bands. We approximated the conversion factor in \textit{F146}-band by averaging the conversion factors in the 2MASS $J$-, $H$-, and $K_S$-bands. 

\begin{figure}
    \centering
    \includegraphics[width=\columnwidth]{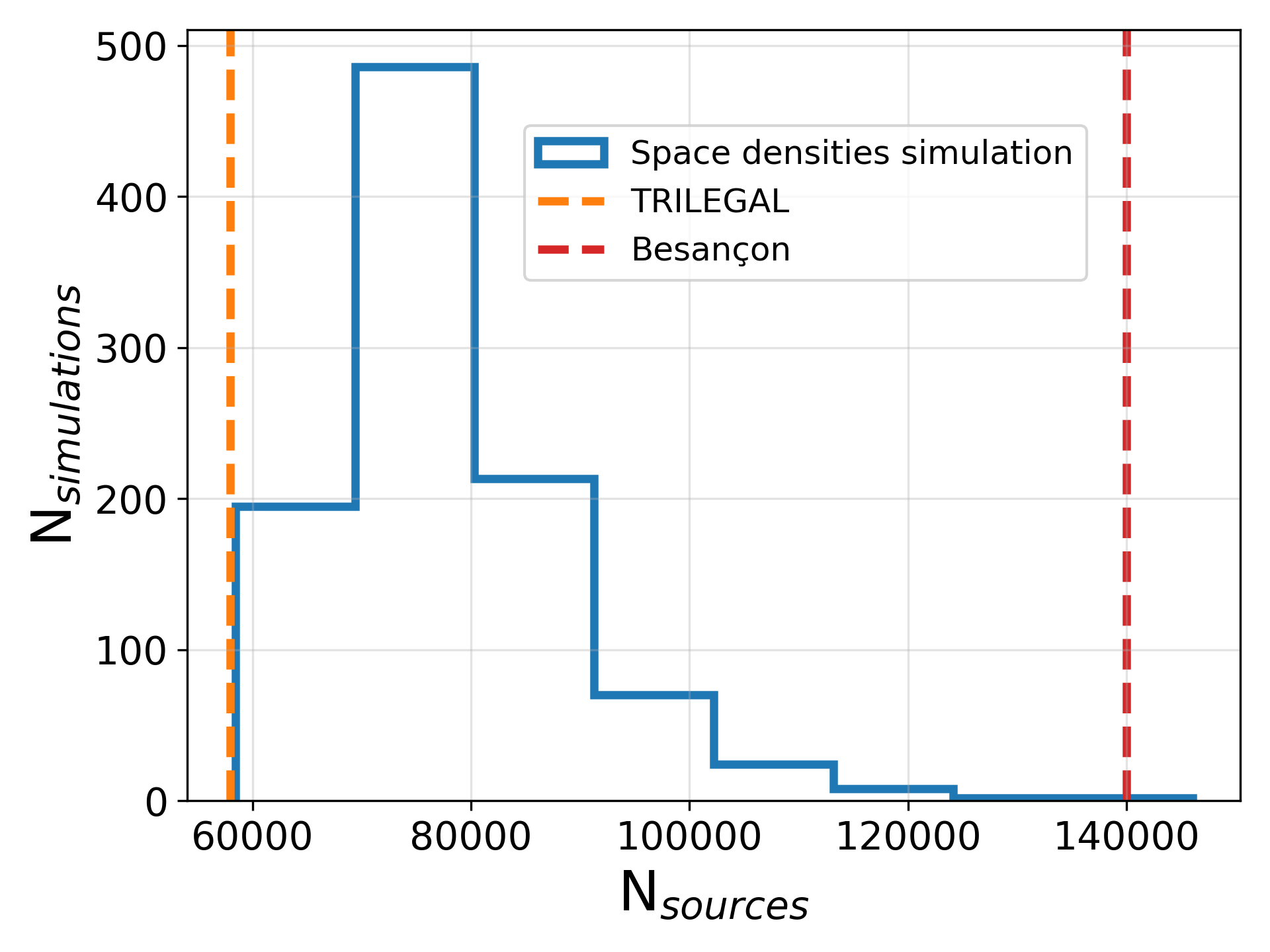}
    \caption{The distribution of the total number of single mid-M and UCD sources with \Fmag $<$ 21 in the survey fields over 1,000 iterations of our simulation with scaled space densities is shown in blue. This simulation predicts an average of 75,500$^{+11,800}_{-7,000}$ single mid-M dwarf and UCD sources in the survey fields (1$\sigma$ range). Predictions from the Galactic stellar population synthesis models \texttt{TRILEGAL} and \texttt{Besançon} are marked with orange and red vertical lines, respectively (see Section \ref{subsubsec:besancon} for more details).}
    \label{fig:n_sources_histogram}
\end{figure}

We repeated this simulation 1,000 times to sufficiently sample the range of source counts that result from randomly sampling the W21 and K21 space densities, $\ell_t$, $\ell_T$, and $f_p$ from Gaussian distributions within each shell. In Figure \ref{fig:n_sources_histogram}, we show distribution of counts of mid-M dwarf and UCD sources with \Fmag~$<$~21, which has a mean of 75,500$^{+11,800}_{-7,000}$ sources.

We show distributions of the average values of various parameters for all sources across the 1,000 instances of our simulation in Figure \ref{fig:source_histograms}. This figure shows magnitudes in the \textit{F146}-, \textit{r}-, \textit{i}-, \textit{z}-, and \textit{J}-bands, along with distances, effective temperatures, spectral types, masses, and radii. The central value and $1\sigma$ range is indicated for each of these distributions. The 99.9\% percentiles on the \textit{r}, \textit{i}, \textit{z}, and \textit{J} magnitudes of our sources are $27.5$, $24.9$, $23.3$, and $21.4$, respectively. Imaging with sensitivity to these magnitude limits (or comparable limits in other red-optical and NIR bands) will thus be required to identify the $\sim$75,500 mid-M dwarf and UCD sources that we expect with \textit{F146}~$<$~21 within the Roman fields. This prospect may be realized with the upcoming Vera C. Rubin Observatory, which will provide seeing-limited imagery of the survey fields with 5$\sigma$ point source depths that will approach or surpass the required magnitude limits over the course of the anticipated 10-year survey operation \citep{Ivezic2019}, or with the Roman survey data itself.

Figure \ref{fig:source_histograms} shows that targets in our simulated sample are located at an average distance of $1963^{+543}_{-667}$ pc. They have masses of $0.18^{+0.03}_{-0.04}$~M$_\odot$, radii of $0.20^{+0.02}_{-0.04}$~R$_\odot$, and spectral types of M4 plus or minus one sub-type, on average. We expect relatively few L and T dwarfs in the Roman fields with \textit{F146}~$<$~21, finding just 57 such sources in each simulation, on average. This is because early-to-mid L dwarfs have the lowest local space densities of any objects considered in our simulation (see Figure \ref{fig:space_densities}), and the small radii and low temperatures of objects in the L and T spectral classes generally place them beyond our \textit{F146}~=~21 cutoff. 

\subsubsection{Note on Metallicity Effects}
\label{subsubsec:metallicity}
The average metallicity of stars in the thin disk increases interior to the Sun's orbit. The furthest sources under consideration in this study, at a distance of 3.5 kpc towards Galactic center, should be slightly enhanced in metals on average, with [Fe/H]~$\approx$~0.3 \citep[e.g.,][]{Andrievsky2002, Pedicelli2009, Genovali2014}. We did not account for the metallicity gradient in our modeling, but we do not expect that its inclusion would significantly impact our results. For fixed NIR luminosity (e.g. $M_K$), the assumed metallicity has a negligible effect on the estimated radius and mass of an M dwarf \citep{Delfosse2000, Bonfils2005, Mann2015, Mann2019}, so including metallicity does not strongly impact the signal-to-noise ratio (S/N) of transit events in our simulated data. Additionally, while there is an observed correlation between host metallicity and the occurrence rate of giant planets around various spectral types \citep{Santos2004, Fischer2005, Johnson2010}, such planets are rarely observed to transit around M-type stars \citep{Dressing2013,Dressing2015,Gaidos2016}. While \citet{Anderson2021} found evidence of higher occurrence rates for ``compact multiple" rocky planet systems around metal-poor K and M dwarfs compared to higher metallicity stars, previous studies found no metallicity dependence for small planet occurrence rates around these spectral types \citep[e.g.,][]{Mann2013, Gaidos2016}. We therefore also ignore the potential effects that enhanced metallicity may have on small planet occurrence rates.


\subsection{Simulated Host Population from Galactic Stellar Population Synthesis Models}
\label{subsubsec:besancon}
In this section, we compare the simulated  counts of mid-M dwarfs and UCDs from Section \ref{subsec:hosts} to predictions from publicly available Galactic stellar population synthesis. Population synthesis models impose IMFs, star formation rates, and stellar evolutionary models to simulate populations of stars in different Galactic components, namely the thin disk, thick disk, halo, and bulge. This is a fundamentally different approach to our method of scaling local volume-complete samples of mid-M dwarfs and UCDs, and can act as an independent check on our predictions. We used the \texttt{TRILEGAL}\footnote{\url{http://stev.oapd.inaf.it/cgi-bin/trilegal}} \citep{Girardi2005} and \texttt{Besançon}\footnote{\url{https://model.obs-besancon.fr/}} \citep{Robin2003, Robin2012, Czekaj2014} models to make these comparisons.

We performed a \texttt{TRILEGAL} simulation in the direction of the center of survey fields ($\ell = 1.0^\circ$, $b = -0.667^\circ$) with a Roman \textit{F146} magnitude limit of 21. We performed this simulation over a field area of 0.1~deg$^2$ due to server limitations on maximum computing time. We assumed the Sun's distance to Galactic center to be 8000~pc. We used the model's default exponential dust extinction law and \citet{Chabrier2001} log-normal IMF with binaries turned off. We used an exponential thin disk with a radial scale length of 2500~pc and a thick disk with a scale length of 2000~pc (i.e., the central values that were used to simulate radial scale lengths in Section \ref{subsec:hosts}). We did not model the halo or bulge populations. This simulation produced 2,956 single sources with $T_{eff}~<~3300$ K over the 0.1 deg$^2$ field area, which equates to about 58,000 sources over the full 1.96 deg$^2$ Roman survey area. This prediction is lower than the average number of sources produced in our simulation in Section \ref{subsec:hosts} by about a factor of 1.3, and it is indicated as a vertical orange line in Figure \ref{fig:n_sources_histogram}. 

\texttt{Besançon} does not currently support the Roman photometric system. Instead, we performed a simulation with magnitude limits based on the 99.9\% percentiles on the \textit{r}-, \textit{i}-, \textit{z}-, and \textit{J}-band magnitudes described in Section \ref{subsec:hosts}. We performed this simulation centered on ($\ell = 1.0^\circ$, $b = -0.667^\circ$) over a field area of 1.96~deg$^2$. The simulation only supports source modeling out to spectral type M9, but such sources dominate the counts of objects with \textit{F146}~$<$~21 in our simulation, so the exclusion of later types will not strongly affect the comparison with our results. 

The \texttt{Besançon} results includes binary systems, with the binary probability as a function of source mass $M$ (in solar masses) given by the following equation from \citet{Arenou2011}:

\begin{equation}
     f(M) = 0.8388 \tanh{0.688M + 0.079} 
\end{equation}

\noindent To approximate the number of single mid-M dwarfs and UCDs, we evaluated this equation for each source with $T_{eff}$~$<$~3300~K, removing it from the list if a number drawn from a random uniform distribution was less than the binary probability determined by its mass. Doing this, we found a total of 140,000 sources with $T_{eff} < 3300$ K in the \texttt{Besançon} simulation, which is a factor of 1.9 times higher than the average predicted by our scaled space densities simulation. This prediction is indicated as a red line in Figure \ref{fig:n_sources_histogram}.

Thus, we find that the average number of mid-M dwarf and UCD sources in the Roman survey fields with \textit{F146}~$<$~21 from our scaled space density simulation lies between the predictions from \texttt{TRILEGAL} and \texttt{Bensaçon}. Furthermore, all three predictions lie within a factor of three of each other. In the remainder of this work, we use the source samples that were created using scaled space densities in Section \ref{subsec:hosts}, as they are intermediate between the predictions from the Galactic stellar population synthesis models. However, we note that if the population of mid-M dwarfs and UCDs in the Roman survey actually resembles the \texttt{TRILEGAL} prediction, our average planet yield estimates would be lower by a factor of about 1.3, whereas if it matches the \texttt{Besançon} prediction, our yield estimates would be enhanced by a factor of about 1.9.

\section{Planet Injection and Recovery Simulation}
\label{sec:injection_recovery}

\subsection{Simulated Planet Population}
\label{subsec:planets}
We simulated planetary systems around each of the 1,000 samples of mid-M dwarfs and UCDs created with our scaled space density simulation in Section \ref{subsec:hosts}. Planets were simulated using the results of \citet{Dressing2015} (hereafter DC15), who used Kepler discoveries to measure planet occurrence rates around early-M dwarfs over a grid of planet radii (0.5$\textendash$4.0 $R_\oplus$) and orbital periods (0.5 $\textendash$ 200 days). These are the nearest spectral types to the ones under consideration in this study with detailed planet occurrence rates. Significant deviations from the number of expected detections under the assumption of the DC15 occurrence rates will test for significant changes in the planet populations of mid-M dwarf and UCD sources (see Section \ref{subsec:scaled_yield}).

For each simulated source, we first applied a random on-sky inclination, drawing from a $\sin{i}$ distribution. We then looped over the DC15 occurrence rate grid, simulating an orbiting planet with a random radius and period within the cell bounds if a randomly chosen number from a uniform distribution was less than the measured occurrence rate in the given cell. In multi-planet systems, we assumed that all planets were co-planar. We tested the resulting planetary system for orbital stability following the approach of \citet{Fabrycky2014}, assuming circular orbits for all planets. For each pair of adjacent planets, we calculated the mutual Hill radius, given by 

\begin{equation}
    R_H = \left[ \frac{M_{in}+M_{out}}{3M_h} \right]^{1/3} \frac{a_{in}+a_{out}}{2}
\end{equation}

where ``in" indicates the inner planet, ``out" indicates the outer planet, $M$ are the planet masses $a$ are the planet semi-major axes, and $M_h$ is the host mass. The planet masses were estimated using the mass-radius-relation from \texttt{Forecaster} \citep{Chen2017}. While a variety of exoplanet mass-radius relationships exist in the literature \citep[e.g.,][]{Weiss2013,Wolfgang2016,Bashi2017,Ning2018,UlmerMoll2019,Unterborn2023}, the Hill radius is weakly dependent on planetary mass, so adopting a different model would likely not alter our stability analysis significantly. If the orbital separation of a pair of planets in units of their mutual Hill radii, $\Delta = (a_{in}-a_{out})/R_H$, is greater than a critical separation of $2\sqrt{3}$, the pair is considered to be stable. We also enforced the conservative heuristic of \citet{Fabrycky2014} for systems with more than three planets, which requires that adjacent inner and outer pairs of planets have a total separation greater than 18 in units of their mutual Hill radii, that is $\Delta_{in} + \Delta_{out} > 18$. We repeated the planet injection procedure for each source until these stability criteria were satisfied. The distribution of planets-per-star created with this procedure is approximately Poissonian with a mean of $2.4$, consistent with the mean of $2.5\pm0.2$ planets-per-star measured for early-M dwarfs in DC15.

We then created light curve models for each planet in the system using \texttt{BATMAN} \citep{Kreidberg2015}, assuming a quadratic limb darkening law with coefficients interpolated from the tables of \citet{Claret2011}. Any planets that were found to transit (including grazing transits) were injected into their host's light curve. In multi-planet systems, each planet was injected individually, ignoring effects from simultaneous transit events. On average, $2656^{+502}_{-266}$ transiting planets were created each simulation.


\subsection{Planet Recovery}
\label{subsec:recovery}
Searching for planets with established transit search algorithms like Box Least Squares \citep{Kovacs2002} or Transit Least Squares \citep{Hippke2019} would be computationally expensive for an average of 75,500 sources each with 40,000-point light curves across 1,000 simulations. Computation time could be sped up by applying these search algorithms to binned data, but the 15-minute cadence of the survey data means that only a handful of points can be averaged together before injected transits are shallowed out significantly. As a result, we adopted the approach of \citet{Montet2017}, and phase-folded the planet-injected light curves on the known orbital period, considering a planet detected if its transit S/N exceeded a value of 7.1. The choice of this significance threshold is motivated by a desire to compare our results with the number of planets around mid-M and UCD spectral types in the Kepler and TESS missions, which used the same threshold. It also permits a direct comparison to the results of \citet{Montet2017}. 

\begin{figure}
    \centering
    \includegraphics[width=\columnwidth]{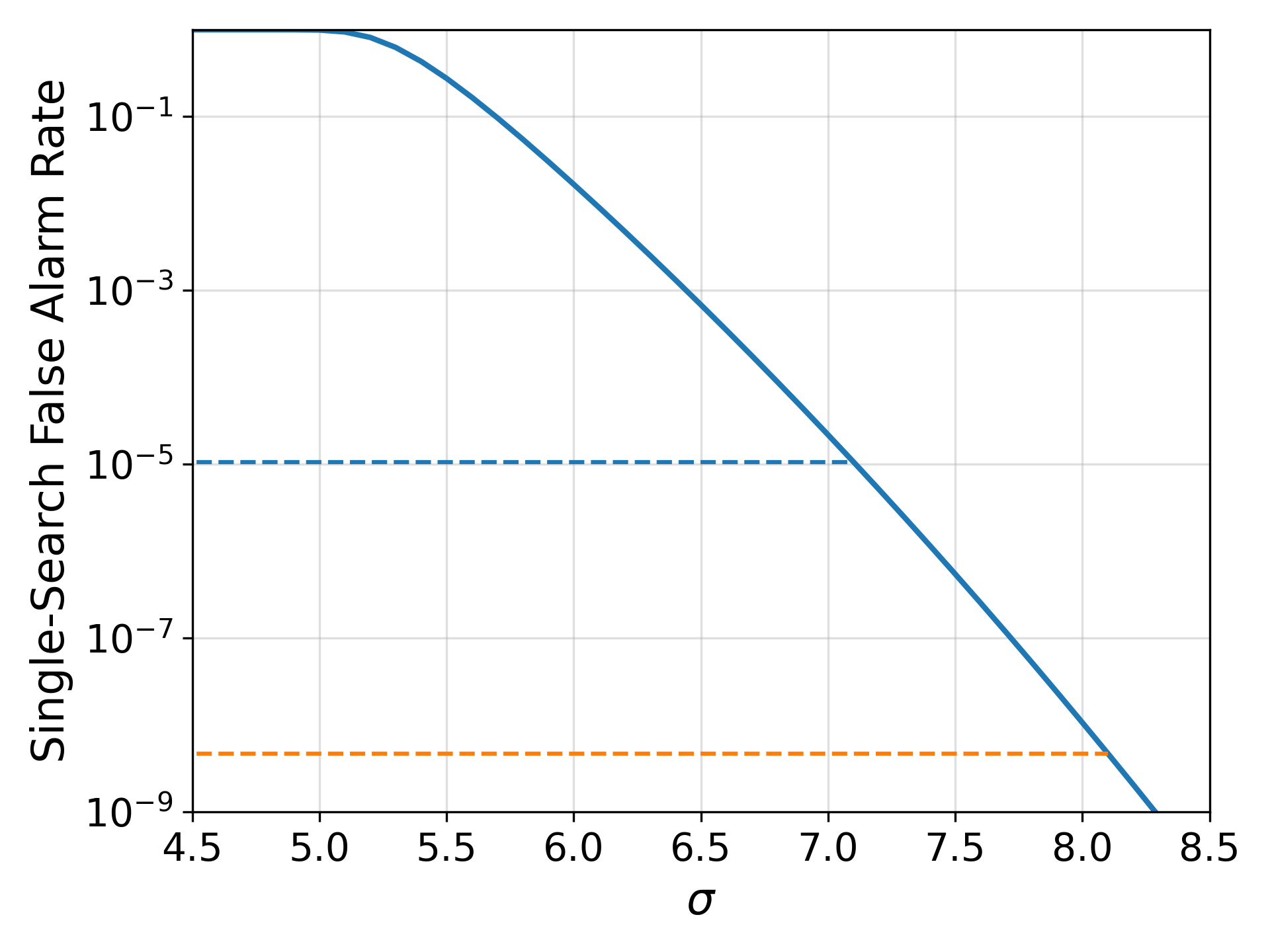}
    \caption{The theoretical single-search false alarm rate as a function of detection threshold for the Kepler mission from \citet{Jenkins2002} (solid blue line). A dashed blue line indicates the desired false positive rate of 1/100,000 for Kepler, which is achieved at a detection threshold of 7.1$\sigma$. A dashed orange line shows an estimate for the Roman survey, where a false positive rate of 1/200,000,000 is achieved at a detection threshold of 8.1$\sigma$.}
    \label{fig:ccdf}
\end{figure}

However, the 7.1$\sigma$ threshold was selected for the Kepler and TESS missions in an attempt to limit the number of statistical false-positive transit detection over the course of the surveys to approximately one. It may not be appropriate for the Roman survey, which will search for transit events around a much larger number of stars \citep[$\sim$200,000,000 versus $\sim$100,000 in the Kepler survey;][]{Spergel2015}. We illustrate this with a simple example in Figure \ref{fig:ccdf}, which shows the theoretical single-search false alarm rate as a function of detection threshold for the Kepler mission from \citet{Jenkins2002}. If we assume that the number of independent statistical tests conducted in searching for transits in a single Roman light curve is roughly equal to the number required for a single Kepler light curve, we can use the theoretical curve to estimate the detection threshold where the false alarm rate equals 1/200,000,000, which it does at 8.1$\sigma$. While a full investigation into the appropriate detection threshold for the Roman survey is beyond the scope of this work, this example illustrates that a higher detection threshold higher than 7.1$\sigma$ may be required to limit the number of statistical false positive detections to one.

We adopt the following definition of transit S/N:

\begin{equation}
    S/N = \frac{\phi_{oot}-\phi_{it}}{\sqrt{\sigma_{oot}^2+\sigma_{it}^2}}
    \label{eq:S/N}
\end{equation}

where $\phi_{oot}$ is the mean out-of-transit flux, $\phi_{it}$ is the mean in-transit flux, $\sigma_{oot}$ is the standard deviation of out-of-transit flux, and $\sigma_{it}$ is the standard deviation of in-transit flux. All of these quantities are evaluated using unbinned data that has been phased on the known orbital period. 

\begin{figure*}
    \centering
    \includegraphics[width=\textwidth]{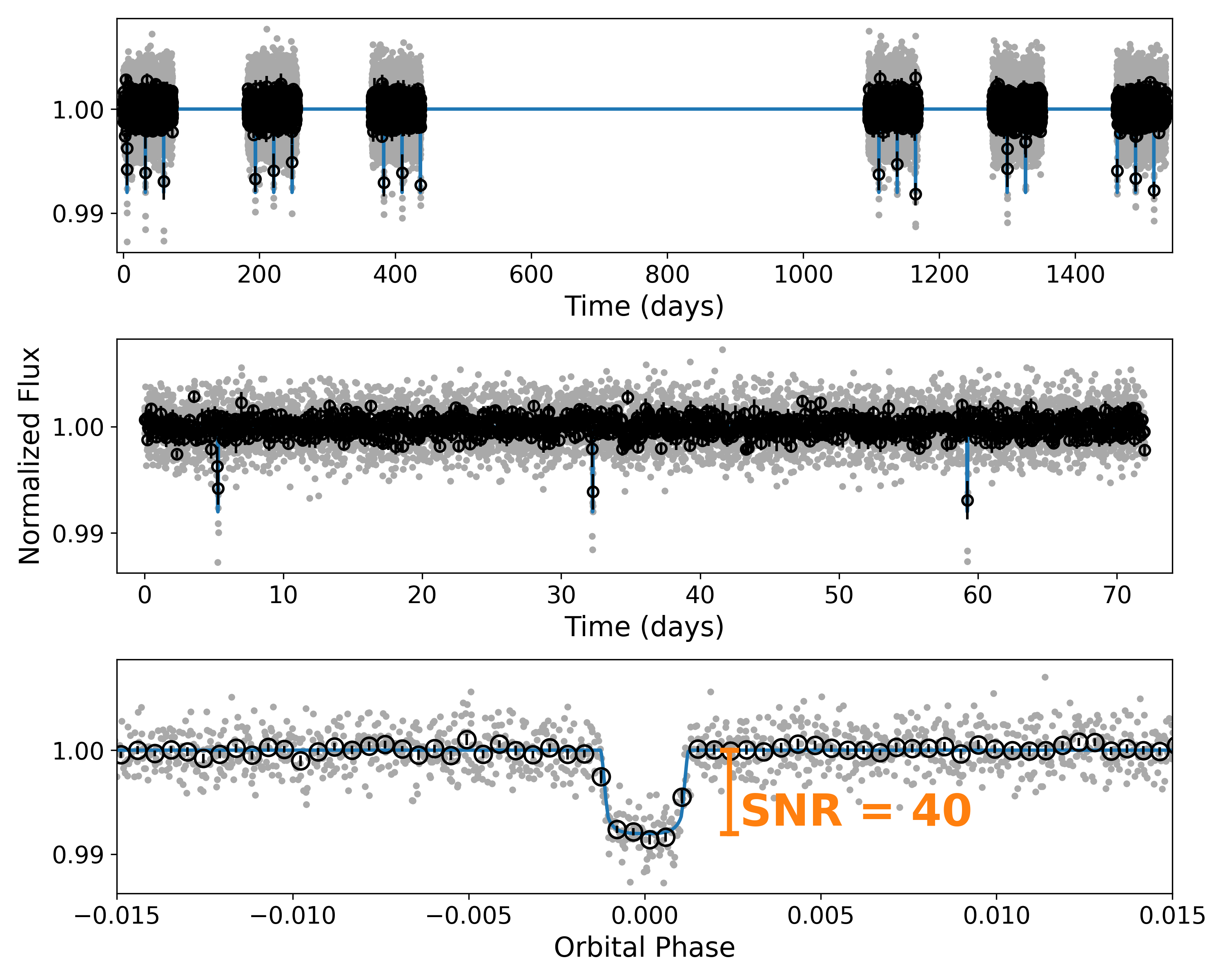}
    \caption{An example detection of a planet in our injection and recovery simulation. The host is an M5 dwarf with \textit{F146}~=~16.5, R$_h$~=~0.133~R$_\odot$, M$_h$~=~0.108~M$_\odot$, T$_{eff}$~=~2895~K, a distance of 193~pc, and an age of 2.68~Gyr. The planet has a radius of 1.24~R$_\oplus$, an orbital period of 27.0~days, $a/R_h$~=~136, and i~=~89.9$^\circ$. Top: the full simulated light curve which consists of six 72-day campaigns, three near the start of the mission and three near the end. Grey points show the 46.8-s photometry evaluated at a 909.6-s cadence, and black points with error bars show the photometry binned over 1.5-hour timescales. The transit model is shown in blue. Middle: The first of the six campaigns. Bottom: the phase-folded data which we use to assess the transit significance. Black points with error bars show the phased photometry binned over 20 points. The transit was recovered in the phased photometry to an S/N of 40 using Equation \ref{eq:S/N}.}
    \label{fig:example_detection}
\end{figure*}

An example planet recovery in our simulation is shown in Figure \ref{fig:example_detection}. It shows simulated Roman time domain survey photometry of an M5 planet host with \textit{F146}~=~16.5, R$_h$~=~0.133~R$_\odot$, M$_h$~=~0.108~M$_\odot$, T$_{eff}$~=~2895~K, a distance of 193~pc, and an age of 2.68~Gyr\footnote{We use the subscript ``$h$" to denote parameters of the planet hosts in our simulation, instead of the traditional ``$s$" for ``star". This is because the sources in our simulation include brown dwarfs.}. A planet was injected into this light curve with a radius of 1.24 R$_\oplus$, an orbital period of 27.0 days, $a/R_h$~=~136, and i~=~89.9$^\circ$. This planet was recovered to an S/N of 40 using Equation \ref{eq:S/N} evaluated on the unbinned phased data.

\section{Results}
\label{sec:results}

\subsection{Total Transiting Planet Yield}
\label{subsec:yield}

We injected planets into our 1,000 samples of mid-M dwarfs and UCD targets. An average of 1347$^{+208}_{-124}$ planets were detected in each simulation under the assumption of DC15 planet occurrence rates. Hosts were frequently found to host 2+ transiting planets, with with an average of 274$^{+41}_{-30}$ multiple systems identified per simulation. Of the detected multiple systems, 78\% were found to be 2-planet systems, 18\% to be 3-planet systems, and 3\% to be 4-planet systems. Systems with five or more of planets were rare, contributing to fewer than half a percent of the total yield of planets detected in multiple systems, combined. The highest number of planets identified in one system across all our simulations was six.

\begin{figure}
    \centering
    \includegraphics[width=\columnwidth]{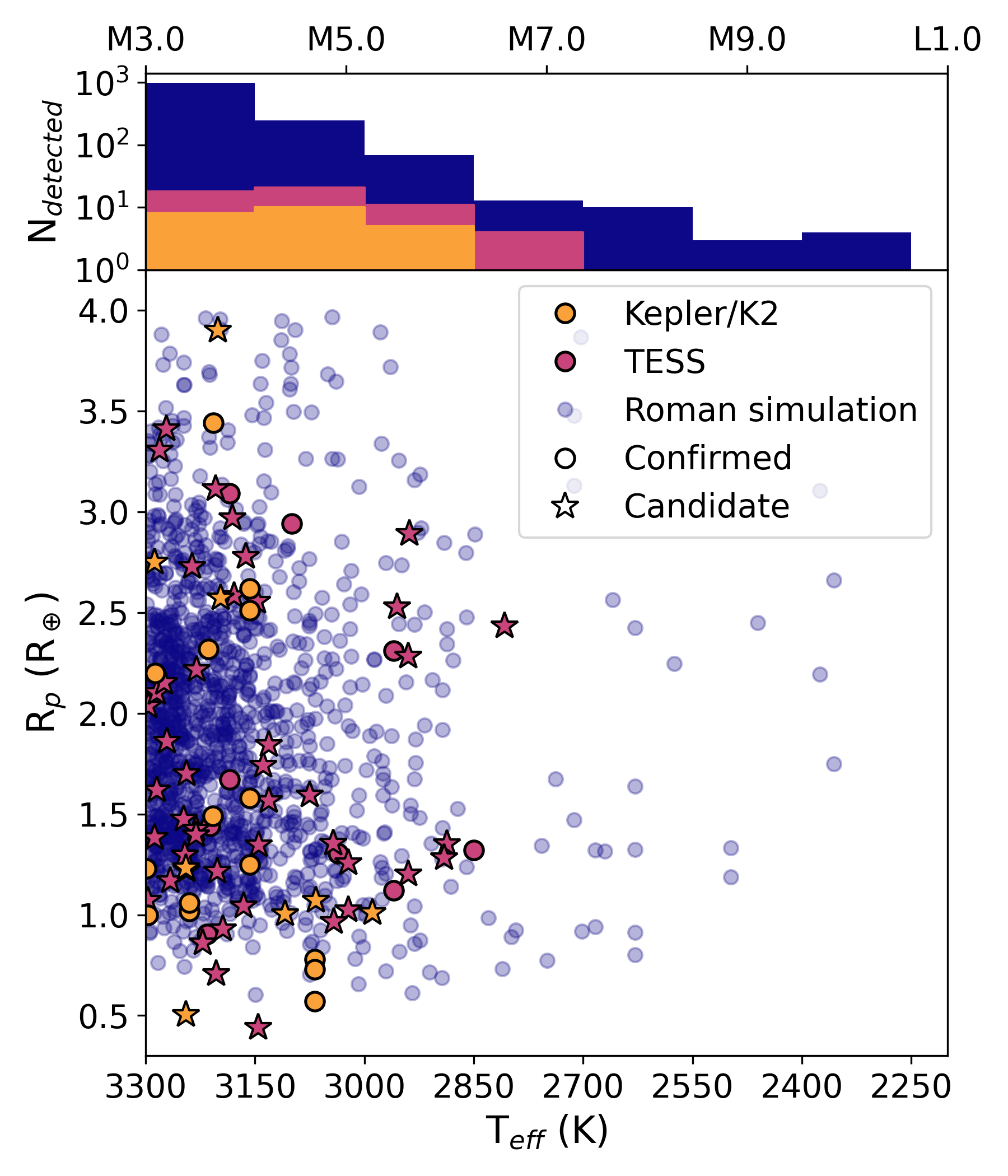}
    \caption{Radii of confirmed (circles) and candidate planets (stars) smaller than 4~$R_\oplus$ versus effective temperature for hosts with T$_{eff}$~$<$~3300~K from the Kepler/K2 (orange) and TESS (pink) missions. Detections from a representative iteration of our simulation (with 1347 total detections) are shown in blue. Effective temperature is listed on the bottom axis while approximate spectral types are given along the top axis. The top panel shows histograms of the number of planet detections as a function of effective temperature for the different missions.}
    \label{fig:kepler_tess_comp}
\end{figure}

We break down the average total number of planets as a function of T$_{eff}$ in Table \ref{tab:host_teffs}. These results are compared against the confirmed and candidate planets with $R_p$~$<$~4~R$_\oplus$ around hosts with T$_{eff}$~$<$~3300~K detected by the Kepler (including K2) and TESS missions, which were tabulated using data from the NASA Exoplanet Archive\footnote{\url{https://exoplanetarchive.ipac.caltech.edu/}} on 2022 October 27. We visualize these results in Figure \ref{fig:kepler_tess_comp}, which shows the planet detections from a representative iteration of our simulation compared to the detections from Kepler and TESS. 

Our results suggest that the Roman survey will provide more than an order of magnitude increase in the number of known small planets around hosts with T$_{eff}$~$<$~3300~K. For example, 77 confirmed or candidate planets with $R_p$~$<$~4~R$_\oplus$ have been identified in Kepler or TESS data around hosts with 2700~K~$<$~T$_{eff}$~$<$~3300 K (SpT $\sim$M3\textendash M7); we predict that Roman will enable the detection of 1334$^{+209}_{-123}$ such planets. While the number of detections are biased toward the earliest spectral types under consideration in this study, we expect that Roman will detect a few dozen planets around UCDs, spectral types that have no confirmed or candidate planets from Kepler or TESS.

\begin{deluxetable}{rlrrr}
        \tablecaption{Planets detected in our injection and recovery simulation as a function of T$_{eff}$/SpT under the assumption of DC15 occurrence rates. The 15.9\%, 50\%, and 84.1\% percentiles are reported for the number of detected planets in our simulations. Few planets were detected on average around L- and T-type dwarfs, so the total detections around these types are presented as a single range. The table also shows confirmed and candidate detections of planets smaller than 4~$R_\oplus$ around these spectral types from the Kepler and TESS missions.}
         \label{tab:host_teffs}
            \tablehead{
                 \colhead{T$_{eff}$} & 
                 \colhead{SpT} & 
                 \colhead{Roman} & 
                 \colhead{Kepler} &
                 \colhead{TESS} 
                 }
            \startdata
                3300$\textendash$3150 & M3.0$\textendash$M4.0 & 488$^{+114}_{-80}$ & 8 & 18\\
                3150$\textendash$3000 & M4.0$\textendash$M5.0 & 631$^{+104}_{-92}$ & 10 & 21\\
                3000$\textendash$2850 & M5.0$\textendash$M6.0 & 163$^{+33}_{-27}$ & 5 & 11\\
                2850$\textendash$2700 & M6.0$\textendash$M7.0 & 47$\pm$11  & 0 & 4\\
                2700$\textendash$2550 & M7.0$\textendash$M8.0 & 18$^{+7}_{-6}$& 0 & 0\\
                2550$\textendash$2400 & M8.0$\textendash$M9.0 & 7$\pm$4 & 0 & 0\\
                2400$\textendash$2250 & M9.0$\textendash$L0.0 & 2$^{+3}_{-1}$ & 0 & 0\\
                2250$\textendash$450 & L0.0$\textendash$T9.0 & 1$^{+1}_{-1}$ & 0 & 0\\
            \enddata
    
            \tablecomments{The reported Kepler detections include confirmed and candidate planets from both the Kepler and K2 missions, excluding candidates that were flagged as false positives. The reported TESS detections include confirmed planets and TESS objects of interest that have not been classified as false positives. The tables detailing these detections were accessed from the NASA Exoplanet Archive on 2022~October~27.}
    \end{deluxetable}

Very few planets were detected around L- or T-type dwarfs in our simulations, with an average of $1^{+1}_{-1}$ detections per simulation. We expect that the search for transiting planets around brown dwarfs will remain the purview of dedicated near-infrared ground-based surveys like SPECULOOS \citep{Delrez2018, Murray2020} and PINES \citep{Tamburo2022a} for the foreseeable future. However, the proposed TEMPO survey of the Orion Nebular Cluster with Roman will be capable of detecting dozens of planets around brown dwarfs and free-floating planetary-mass objects, because the proximity ($\sim$400 pc) and young ages ($\sim$1-3 Myr) of these sources  make them much brighter on average than the sources in the Galactic bulge time domain survey \citep{Limbach2022}. 

\begin{figure*}
    \centering
    \includegraphics[width=\textwidth]{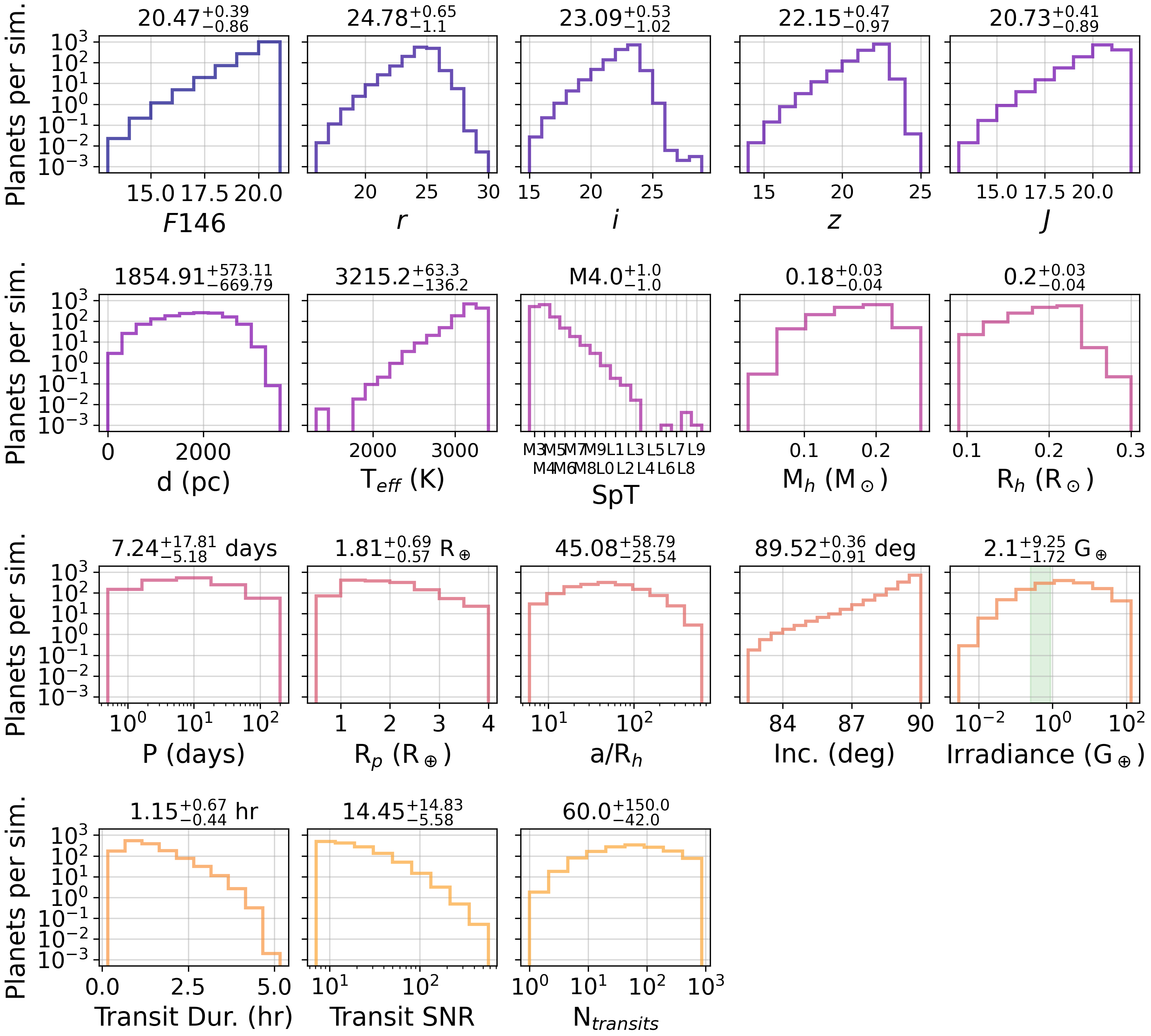}
    \caption{Properties of transiting planet hosts and transiting planets detected across our 1,000 simulations. As in Figure \ref{fig:source_histograms}, the distributions are shown on a log scale and were normalized to show the average counts in a single simulation. The 15.9\%, 50\%, and 84.1\%  percentiles are reported at the top of each panel. First row: \textit{F146}-, \textit{r}-, \textit{i}-, \textit{z}-, and \textit{J}-band magnitudes of hosts with detected transiting planets. Second row: distance in pc, $T_{eff}$ in K, spectral types, masses in $M_\odot$, and radii in $R_\odot$ of hosts with detected transiting planets. Third row: orbital periods, radii in R$_\oplus$, $a/R_h$, inclinations in degrees, and irradiances in Earth irradiance ($G_\oplus$) of detected transiting planets. The irradiance distribution includes a shaded area indicating the conservative HZ of \citet{Kopparapu2013}. Fourth row: durations in hours, S/N, and number of transits of detected transiting planets.}
    \label{fig:planet_properties_histograms}
\end{figure*}

We show distributions of detected planet properties in Figure \ref{fig:planet_properties_histograms}. The average planet had an orbital period of $7.2^{+17.8}_{-5.2}$ days, a radius of $1.8^{+0.7}_{-0.6}$ R$_\oplus$, an $a/R_h$ of $45.1^{+58.8}_{-25.5}$, an irradiance of $2.1^{+9.3}_{-1.7}$ G$_\oplus$, a transit duration of $1.2^{+0.7}_{-0.4}$ hours, and was detected to an S/N of $14.5^{+14.8}_{-5.6}$ with an average of $60^{+150}_{-42}$ transits.

    As discussed in Section \ref{subsec:recovery}, we use the transit S/N of a target's phase-folded light curve to determine whether an injected planet is recovered or not; we impose no constraints on the minimum number of transits for a planet to be considered detected. Our detections therefore include some planets with fewer than three transits, which would require further transit detections to measure their true orbital periods. However, since the maximum orbital periods of our injected planets are 200~days, and since the simulated light curves consist of 432~days of data, such detections are rare. We find just 0.02\% of our detected planets exhibit single transits in the light curves, while 0.11\% transit twice. 

    As noted in Section~\ref{subsec:photometry}, our photometric model assumes a constant noise floor of 1~mmag, which becomes the dominant noise source at magnitudes brighter than \Fmag~$\approx$~16.8. Only 0.3\% of the planets in our simulation were detected around sources with \Fmag~$<$~16.8, indicating that the assumed noise floor has negligible impact on the total planet yield.

    Finally, as discussed in Section \ref{subsec:recovery}, a detection threshold of $\sim$8.1$\sigma$ may be appropriate for the Roman survey in order to limit the number of statistical false-positive transit detections to one. On average, we find that 91\% of our detections have S/N  values greater than 8.1, meaning that our total transit yield would be lower by 9\% if the more conservative detection threshold was adopted.

\subsection{Habitable Terrestrial Transiting Planet Yield}
\label{subsec:hz_planets}
Terrestrial planets around mid-to-late M dwarfs enable the atmospheric characterization of habitable Earth-sized planets with current and near-future telescope facilities \citep[e.g.,][]{Morley2017}. We performed a simple assessment of the habitability of planets in our simulation by calculating the irradiance of each in Earth irradiance:

\begin{equation}
    \frac{G}{G_\oplus} = \frac{L}{4 \pi a^2 G_\oplus} = \frac{\sigma R_h^2 T_{eff}^4}{a^2 G_\oplus}
\end{equation}

\noindent where $\sigma$ is the Stefan–Boltzmann constant, $R_h$ is the radius of the host, $T_{eff}$ is the effective temperature of the host, $a$ is the semi-major axis of the planet's orbit, and $G_\oplus$ is the Earth's average annual irradiance at the top of the atmosphere, which is about 1361~$W/m^2$. We considered a planet to be ``habitable'' if it experienced an irradiance between 0.25$\textendash$0.88 G$_\oplus$, the limits of the conservative maximum greenhouse/moist greenhouse habitable zone (HZ) irradiance limits of \citet{Kopparapu2013}.

Under the assumed DC15 occurrence rates, we predict an average of 37$^{+8}_{-7}$ habitable planets with radii less than $1.48$ R$_\oplus$, the boundary between rocky and non-rocky worlds identified by \citet{Rogers2015}. These planets are detected around stars with an average \textit{F146} magnitude of 20.1$^{+0.6}_{-1.1}$. If we instead assume the smaller rocky/non-rocky planet transition of $1.23$~R$_\oplus$ identified by \citet{Chen2017}, we predict an average of 13$^{+4}_{-3}$ habitable terrestrial planet detections. These planets, being smaller, are preferentially detected around slightly brighter stars, with an average \textit{F146} magnitude of 19.9$^{+0.8}_{-1.3}$. However, it should be noted that it is difficult to assess whether or not a planet is rocky on the basis of its radius alone due to degeneracies with mass and composition. For example, \citet{Unterborn2023} showed that there exist regions of radius-mass-composition space in which planets are ``nominally rocky" up to the 2~R$_\oplus$ limit that was considered in their study.

Regardless, it is exceedingly unlikely that the atmospheres of transiting HZ terrestrial planets detected by Roman around mid-M dwarfs and UCDs will be detectable with JWST. Their host stars will be roughly 10 magnitudes fainter on average than the planet hosts for which these measurements are currently feasible. While a full assessment of the feasibility of atmospheric observations of these planets is beyond the scope of this work, their characterization would likely be infeasible even with the next generation of space telescopes like the Large UV/Optical/Infrared Surveyor (LUVOIR), whose 15-m design (LUVOIR-A) will have a factor of $\sim$5$\times$ the light collecting area of JWST \citep{Luvoir2019}.

\subsection{Yield Differences with Scaled Planet Occurrence Rates}
\label{subsec:scaled_yield}

As noted in Section \ref{sec:intro}, \citet{HardegreeUllman2019} (hereafter HU19) measured a tentative increase in the occurrence rates of short-period planets with orbital periods between 0.5$\textendash$10 days and radii between 0.5$\textendash$2.5~$R_\oplus$ around mid-type M dwarfs, reporting $0.86^{+1.32}_{-0.68}$, 
$1.36^{+2.30}_{-1.02}$, and
$3.07^{+5.49}_{-2.49}$ such planets per star over the M3, M4, and M5 spectral types, respectively. DC15 measured $0.62$ such planets around early-M dwarfs, so the HU19 results suggest a progressive increase in the occurrence rates of such planets out to mid-M spectral types (albeit with large error bars). 

We performed a version of our planet injection and recovery simulation with scaled DC15 planet occurrence rates to test whether the Roman time domain survey can reveal the enhancement around mid-M spectral types reported in HU19. Specifically, we again applied the occurrence rate grid of DC15, but scaled their reported occurrence rates for planets with radii between $0.5\textendash2.5$~R$_\oplus$ and periods between $0.5\textendash10$ days. We enhanced the occurrence rates around M3 dwarfs by a factor of $0.86/0.62 = 1.39$, around M4 dwarfs by a factor of $1.36/0.62 = 2.19$, and around M5 dwarfs by a factor of $3.07/0.62 = 4.98$. The occurrence rates of planets around later spectral types or outside of the specified radius and period limits were left unchanged. 

\begin{figure}
    \centering
    \includegraphics[width=\columnwidth]{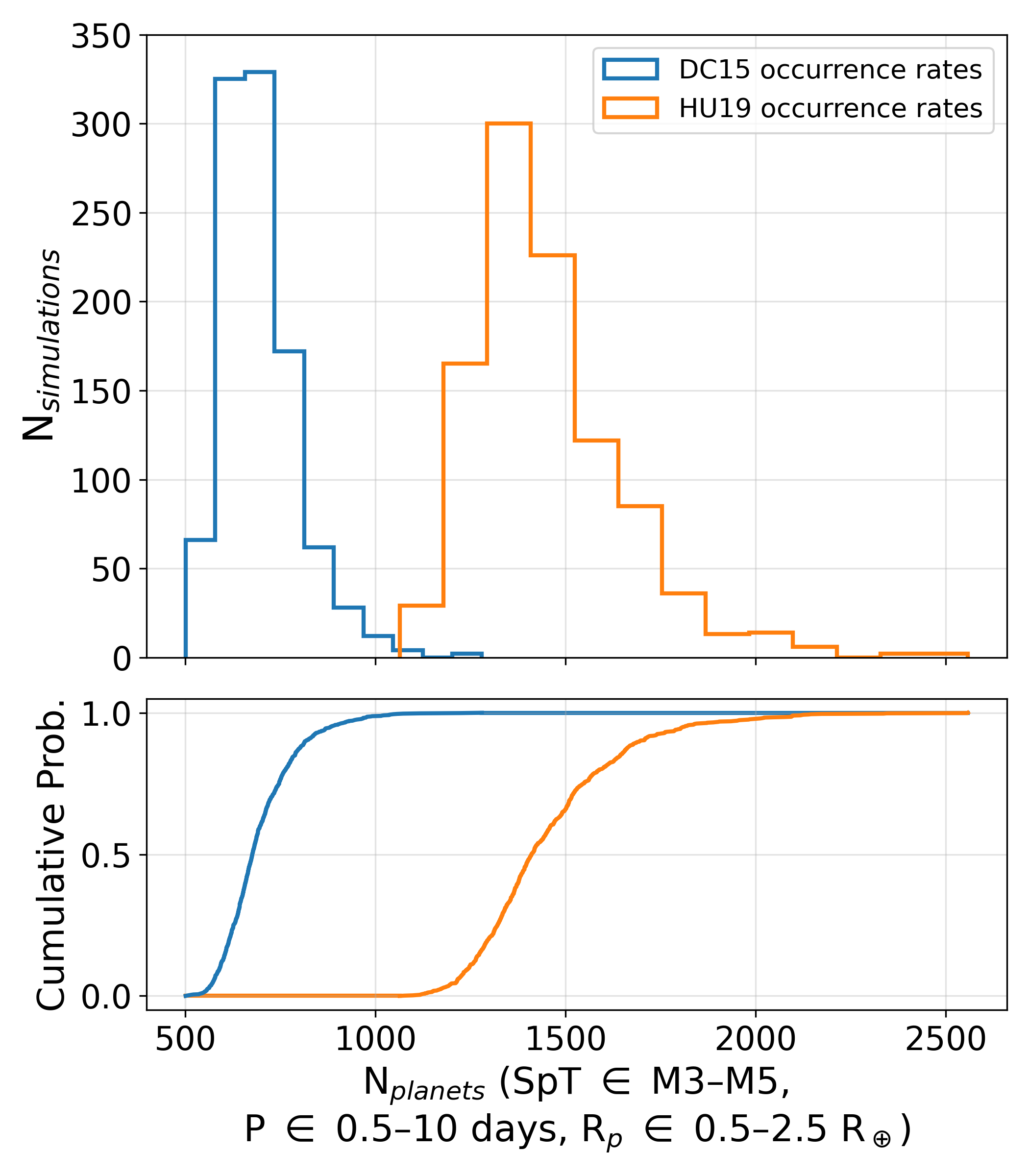}
    \caption{Top: Distributions of detected planets with radii between 0.5$\textendash$2.5 R$_\oplus$ and orbital periods between 0.5$\textendash$10 days around M3$\textendash$M5 dwarfs in our simulations. The results using the DC15 planet occurrence rates for early-M dwarfs are shown in blue, and the results using the HU19 enhanced planet occurrence rates for M3$\textendash$M5 dwarfs are shown in orange. Bottom: Cumulative distribution functions for the two distributions shown in the top panel.}
    \label{fig:planet_yield}
\end{figure}

Figure \ref{fig:planet_yield} shows the effect on planet yield for M3$\textendash$M5 dwarfs, specifically. In the top panel, it shows distributions of transiting planets with radii 0.5$\textendash$2.5~R$_\oplus$ and orbital periods between 0.5$\textendash$10 days around M3$\textendash$M5 dwarfs detected across all 1,000 simulations for planet populations with DC15 and HU19 occurrence rates. The bottom panel shows the cumulative distribution functions of the two distributions. An average of 676$^{+105}_{-69}$ such planets are detected under the assumption of the DC15 occurrence rates, whereas 1411$^{+226}_{-133}$ are detected on average assuming HU19 occurrence rates. These two average predictions are different by 4.3$\sigma$, which suggests that the yield of transiting planets from the Roman Galactic Bulge Time Domain Survey will be able to determine whether planet occurrence rates are enhanced around mid-M dwarf and UCDs compared to early-M dwarfs to high significance. However, this prediction assumes that the enhanced metallicity expected for our sources (up to [Fe/H] $\approx$ 0.3, see Section \ref{subsubsec:metallicity}) does not significantly affect the occurrence rates of short-period planets smaller than 2.5 R$_\oplus$, and this assumption would have to be tested in reality.

\section{Conclusions}
\label{sec:conclusions}
In this paper, we investigated the potential of the Nancy Grace Roman Space Telescope time domain survey of the Galactic bulge for the purposes of detecting transiting exoplanets around mid-M dwarfs and UCDs.

We summarize our results as follows:
\begin{enumerate}
    \item We predict an average of 75,500$^{+11,800}_{-7,000}$ single mid-M dwarfs and UCDs in the Roman time domain survey fields with \textit{F146}~$<$~21, a quantity that lies between the predictions from two stellar population synthesis codes, \texttt{Besançon} and \texttt{TRILEGAL}.

    \item We predict an average of 1347$^{+208}_{-124}$ transiting planet detections around single mid-M dwarf and UCD sources in the Roman Galactic Bulge Time Domain Survey if their planet population matches that measured for early-M dwarfs in DC15. This quantity of detections will permit constraints on planet occurrence rates out to spectral type $\sim$M7. 
    
    \item A couple dozen planets will be detected around spectral types M7\textendash M9, but few (if any) will be detected around L- and T-type dwarfs. The detection of transiting planets around brown dwarfs will thus likely remain the purview of dedicated ground-based surveys like SPECULOOS \citet{Delrez2018,Murray2020} and PINES \citep{Tamburo2022a}, though the proposed TEMPO survey of the Orion Nebular Cluster with Roman \citep{Limbach2022} will be able to detect dozens of planets around young brown dwarfs. The Rubin LSST project may also be capable of probing this parameter space with its Deep Drilling Fields \citep[e.g.,][]{Lund2015,Ivezic2019}, but will be subject to the normal limitations of ground-based observations for making transiting exoplanet discoveries.
    
    \item We predict the detection of 37$^{+8}_{-7}$ planets within the conservative \citet{Kopparapu2013} HZ with radii smaller than 1.48 $R_\oplus$, and 13$^{+4}_{-3}$  with radii smaller than 1.23~$R_\oplus$. These radius limits correspond to the transitions between rocky and gaseous planets identified in \citet{Rogers2015} and \cite{Chen2017}, respectively. However, except in rare cases, we predict that the atmospheres of these planets will not be characterizable with current or near-future space telescope facilities, owing to the faintness of their host stars.

    \item If the occurrence rates of planets smaller than 2.5 R$_\oplus$ on orbital periods less than 10 days are enhanced around M3$\textendash$M5 dwarfs as measured in \citet{HardegreeUllman2019}, the yield of transiting planets from the Galactic bulge time domain survey could reveal that enhancement to a significance of over 4$\sigma$. However, we note that the Roman survey will detect planets around stars with slightly higher average metallicities than the early-M dwarf hosts that were observed with Kepler, so the effects of metallicity will have to be accounted for when determining the significance of an observed enhancement in planet occurrence rates.
    
\end{enumerate}

Our conclusions are subject to several important caveats which should be borne in mind when interpreting our results. For one, we simulated light curves assuming purely Gaussian noise, neglecting effects from systematics (e.g., intrapixel effects) or source variability (e.g., rotation or flares). Contributions from these noise sources would only serve to decrease our expected planet yield. Second, we have ignored the fact that planet occurrence rates for early M dwarfs can be better explained by a dual-population mixture model rather than the single population that we have assumed here \citep{Ballard2016}. Our results, therefore, likely underestimate the number of single-planet systems and overestimate the number of two-planet systems, and potentially underestimate the total transiting planet yield compared to multiple-population models \citep[e.g.,][]{Ballard2019}. Third, we have implicitly assumed that all transiting planets that are detectable to a significance of at least 7.1$\sigma$ in the survey data \textit{will} be detected, and we did not consider statistical or astrophysical false positives. Confirming the planetary nature of all candidate systems found in the survey data will require significant effort and will be complicated by the crowded nature of the survey fields which will have a crowding rate about two times higher than Kepler \citep{Montet2017}. In a related manner, we did not consider the effects of flux contamination in our simulations, which would serve to reduce the S/N of transit events and reduce the total planet yield. Finally, we performed our planet injection and recovery simulation assuming the ``Cycle 7" design of the Roman survey as described in \citet{Penny2019}. The exact implementations of Roman's three Core Community surveys have not yet been finalized, and changes to the design of the Galactic Bulge Time Domain Survey (e.g., number of fields, cadence, filter choices, etc.) would naturally affect our predicted planet yield.

With these caveats in mind, our results have demonstrated that the Nancy Grace Roman Space Telescope Galactic Bulge Time Domain Survey has the potential to transform our understanding of planet occurrence rates around very-low mass stars, optically faint sources that have not yet been probed with sufficiently sensitive data in a large-scale transit survey. These constraints are needed to differentiate between two competing predictions: occurrence rate trends from the Kepler mission, which found a significant anti-correlation between host mass and the occurrence rate of short-period 1$\textendash$4~R$_\oplus$ transiting exoplanets, and planet formation models, which generally predict that the formation of super-Earths and mini-Neptunes should become more difficult around mid-M dwarfs and later. Data from the Roman Galactic Bulge Time Domain Survey will be sensitive enough to place these constraints on spectral types out to $\sim$M7, and will vastly improve our understanding of planetary systems around objects at the bottom of the main sequence. Based on the stellar IMF, mid-M dwarfs and UCDs are some of the most common outcomes of the star formation process; as a result, their planets may comprise the bulk of the galactic census of exoplanets. Roman represents the first mission with the capacity to test this possibility and it could allow us to understand our own Solar System in a galactic context for the first time.

\begin{acknowledgments}

The authors thank their anonymous referee, whose comments improved the quality of this work.

This material is based upon work supported by the National Aeronautics and Space Administration under Grant No. 80NSSC20K0256 issued through the Science Mission Directorate.

This research has made use of the NASA Exoplanet Archive, which is operated by the California Institute of Technology, under contract with the National Aeronautics and Space Administration under the Exoplanet Exploration Program.

The authors thank Jennifer Winters for helpful input regarding the binarity of sources in the W21 sample.
\end{acknowledgments}

\vspace{5mm}
\facilities{ NASA Exoplanet Archive
 }

\software{ \texttt{BATMAN} \citep{Kreidberg2015}, \texttt{Besançon} \citep{Robin2003}, \texttt{dustmaps} \citep{Green2018b}, \texttt{Forecaster} \citep{Chen2017}, \texttt{TRILEGAL} \citep{Girardi2005}}

\bibliography{biblio}
\bibliographystyle{aasjournal}

\end{document}